\documentclass[traditabstract]{aa}  

\usepackage{graphicx}
\usepackage{natbib}
\usepackage{lscape}
\usepackage{longtable}
\usepackage{txfonts}

\usepackage{float}

\def\mstar  {$M_{\star}$}
\def\macc   {$\dot{M}_{\rm acc}$}
\def\lacc   {$L_{\rm acc}$}
\def\mdust {$M_{\rm disk,dust}$}
\def\mdisk {$M_{\rm disk}$}

\def\msun {$M_{\odot}$}
\def\lsun {$L_{\odot}$}
\def\lstar {$L_\star$}
\newcommand{\degree}{\ensuremath{^\circ}}

\def\nodata {...}

\usepackage{color}

\begin{document}

   \title{X-Shooter survey of disk accretion in Upper Scorpius} \subtitle{I. Very high accretion rates at age$>$5 Myr\thanks{Based on observations collected at the European Southern Observatory under ESO programmes 097.C-0378(A) and 0101.C-0866(A).} }

\titlerunning{X-Shooter survey of disk accretion in Upper Scorpius}
\authorrunning{Manara et al.}


   \author{C.F. Manara \inst{1}\fnmsep\thanks{ESO Fellow} \and A. Natta\inst{2} \and G.P. Rosotti \inst{3} \and J.M. Alcal\'a \inst{4} \and B. Nisini \inst{5} \and G. Lodato\inst{6} \and L. Testi\inst{1,7} \and I. Pascucci \inst{8}   \and \\L. Hillenbrand \inst{9} \and J. Carpenter \inst{10} \and A. Scholz \inst{11} \and D. Fedele\inst{7} \and A. Frasca\inst{12} \and G. Mulders \inst{13} \and E. Rigliaco\inst{14} \and \\C. Scardoni\inst{6,15} \and E. Zari \inst{16}
          }

   \institute{European Southern Observatory, Karl-Schwarzschild-Strasse 2, 85748 Garching bei M\"unchen, Germany\\  \email{cmanara@eso.org}
\and 
School of Cosmic Physics, Dublin Institute for Advanced Studies, 31 Fitzwilliams Place, Dublin 2, Ireland
\and
Leiden Observatory, Leiden University, PO Box 9513, 2300 RA Leiden, The Netherlands
\and
INAF -- Osservatorio Astronomico di Capodimonte, via Moiariello 16, 80131 Napoli, Italy
\and
INAF -- Osservatorio Astronomico di Roma, via di Frascati 33, 00078 Monte Porzio Catone, Italy
\and
Dipartimento di Fisica, Universit\'a degli Studi di Milano, Via Giovanni Celoria 16, I-20133 Milano, Italy
\and 
INAF -- Osservatorio Astrofisico di Arcetri, L.go E. Fermi 5, 50125 Firenze, Italy
\and
Lunar and Planetary Laboratory, The University of Arizona, Tucson, AZ 85721, USA
\and
California Institute of Technology, 1200 East California Blvd, Pasadena, CA 91125, USA
\and
Joint ALMA Observatory, Avenida Alonso de C\'ordova 3107, Vitacura, Santiago, Chile
\and
SUPA, School of Physics \& Astronomy, University of St Andrews, North Haugh, St Andrews, KY16 9SS, United Kingdom
\and 
INAF -- Osservatorio Astrofisico di Catania, via S. Sofia, 78, 95123 Catania, Italy
\and
Department of the Geophysical Sciences, The University of Chicago, 5734 South Ellis Avenue, Chicago, IL 60637, USA
\and
INAF -- Osservatorio Astronomico di Padova, Vicolo dell'Osservatorio 5, I-35122, Padova, Italy
\and
Institute of Astronomy, University of Cambridge, Madingley Road, Cambridge CB3 OHA, UK
\and
Max Planck Institute for Astronomy, K\"onigstuhl 17, 69117 Heidelberg, Germany
}

 \date{Received Mar 13, 2020; accepted May 11, 2020}

 
  \abstract
  {Determining the mechanisms that drive the evolution of protoplanetary disks is a necessary step toward understanding how planets form. For this work, we measured the mass accretion rate for young stellar objects with disks at age $>$5 Myr, a critical test for the current models of disk evolution. 
  We present the analysis of the spectra of 36 targets in the $\sim$5-10 Myr old Upper Scorpius star-forming regions for which disk masses were measured with ALMA. 
  We find that the mass accretion rates in this sample of old but still surviving disks are similarly high as those of the younger ($\sim 1-3$ Myr old) star-forming regions of Lupus and Chamaeleon~I, when considering the dependence on stellar and disk mass. In particular, several disks show high mass accretion rates $\gtrsim 10^{-9}M_\odot$/yr while having low disk masses. Furthermore, 
  the median values of the measured mass accretion rates in the disk mass ranges where our sample is complete at a level $\sim 60-80$\% are compatible in these three regions. 
  At the same time, the spread of mass accretion rates at any given disk mass is still $>$0.9 dex, even at age$>$5 Myr. 
  These results are in contrast with simple models of viscous evolution, which would predict that the values of the mass accretion rate diminish with time, and a tighter correlation with disk mass at age$>$5 Myr. 
  Similarly, simple models of internal photoevaporation cannot reproduce the observed mass accretion rates, while external photoevaporation might explain the low disk masses and high accretion rates. 
  A possible partial solution to the discrepancy with the viscous models is that the gas-to-dust ratio of the disks at $\sim$5-10 Myr is significantly different and higher than the canonical 100, as suggested by some dust and gas disk evolution models. The results shown here require the presence of several interplaying processes, such as detailed dust evolution, external photoevaporation, and possibly MHD winds, to explain the secular evolution of protoplanetary disks. }


   \maketitle
%

\section{Introduction}
The study of the evolution of planet-forming disks around young stars and their ability and modality to form planets strongly relies on describing how the main disk properties evolve with time and depend on the properties of the central star. 

From a theoretical viewpoint, the evolution of the disk and its dispersal is commonly described as an interplay between accretion of material through the disk and onto the central star \citep[e.g.,][]{HCH16}, dispersal of material through winds \citep[e.g.,][]{EP17}, and internal processes leading to grain growth and planet formation \citep[e.g.,][]{testi14,MR16}. On top of that, external processes, such as external photoevaporation and dynamical interactions, can also affect the evolution of disks \citep[e.g.,][]{winter18}.

A number of disk properties, such as the mass accretion rate onto the central star (\macc), the mass-loss rate in winds, and the disk mass (\mdisk), can now be measured in a large number of objects in different evolutionary stages. This is made possible thanks to the availability of sensitive optical spectrographs, such as the X-Shooter instrument on the Very Large Telescope (VLT), and millimeter intereferometers, in particular the Atacama Large Millimeter and sub-millimeter Array (ALMA). 

It is the combination of these instruments that allowed us to establish that the disk mass and \macc \ are correlated \citep{manara16b,mulders17}. This relation is predicted by the viscous evolution model \citep[e.g.,][]{LBP74,hartmann98,dullemond06,lodato17,mulders17,rosotti17}. However, the correlation measured in the young star populations of the $\sim$1-3 Myr old Lupus and Chamaeleon~I star-forming regions is in line with the expectations of viscous evolution theory only if the typical viscous timescales have a large spread of values and are typically of the order of the age of the region $\sim$1 Myr \citep{lodato17,mulders17}. Such a long viscous timescale is needed to explain the observed scatter of the relation ($\sim$1 dex), much larger than what is predicted using shorter viscous timescales \citep[e.g.,][]{dullemond06,mulders17,manara19b}. Assuming purely viscous evolution, a tight correlation with a much smaller spread of \macc \ at any \mdisk \ is expected at older ages $>$5 Myr. At this time, the spread in this relation should be dominated by uncertainties on the \macc \ estimates if viscous accretion is the driver of the evolution of disks.

On the other hand, other processes can also affect the ratio between \mdisk{} and \macc{} at different ages. \citet{rosotti17} expanded the work of \citet{jones12} to show that internal processes, such as internal photoevaporation, planet formation, or the presence of dead zones, would make the \macc/\mdisk \ ratio smaller than what is expected by pure viscous evolution. This was recently confirmed by more detailed description of the evolution of \macc \ and \mdisk \ in the case of internal photoevaporation by \citet{somigliana20}. On the contrary, external photoevaporation would remove material from the disk causing an increase of the \macc/\mdisk \ ratio with respect to pure viscous evolution. 

All the aforementioned processes can be critically tested by looking at the \mdisk-\macc \ relation in different samples of young stellar objects at different ages and in different environments. 
Here we present the results of the first survey of accretion rates in the disk-bearing stars of the $\sim$5-10 Myr old \citep{PM16,feiden16,david19} Upper Scorpius star-forming region.
Our initial aim is to establish, for the first time, the nature of the relation between \macc \ and \mdisk \ at ages $>$5 Myr, and, secondly, to provide a measurement of the typical median values of \macc \ and of the scatter of this relation. The empirical constraints on the time evolution of the \macc-\mdisk \  relation will allow us to further constrain how protoplanetary disks evolve. 

The paper is structured as follows. Section~\ref{sect::data} presents the sample selection, observations, and the data reduction procedure. The analysis of the spectra is then presented in Sect.~\ref{sect::analysis}, while the results of our analysis are described in Sect.~\ref{sect::results}. We then discuss our findings in Sect.~\ref{sect::discussion} and outline the conclusions of this work in Sect.~\ref{sect::conclusions}.


\section{Sample, observations, and data reduction}\label{sect::data}

\subsection{Sample}
We selected our sample starting from the ALMA observations by \citet{barenfeld16}, which included all the objects known at the time to have infrared excess, and therefore a disk, and with spectral types from G2 to M4.75 \citep{luhman12,carpenter06}. Additional candidate members of the region were found later on \citep[e.g.,][]{wilkinson18}. Of the 106 targets observed by  \citet{barenfeld16}, we excluded the 31 ``debris/evolved transitional sources'', as they probably represent either young debris disks composed of second-generation dust, or amorphous disks (which are not the targets of this study), as well as the 22 ALMA nondetections of ``primordial'' disks. 
The latter are excluded as their disk masses are lower than those considered in the analysis of this work, as discussed in the following. 
The values of disk dust masses (\mdust) were obtained by \citet{barenfeld16} from the millimeter flux, assuming a disk temperature dependent on the stellar luminosity and a single opacity and distance ($d=145$ pc) for all disks, and with the assumption that the disk thermal emission is optically thin at the wavelength of the observations \citep[0.88 mm,][]{barenfeld16}. We revisit these estimates based on the individual distances obtained from the parallaxes provided by the Gaia data release 2 \citep[DR2,][see Table~\ref{tab::res}]{gaia,gaiadr2}. 

\begin{figure}[]
\centering
\includegraphics[width=0.45\textwidth]{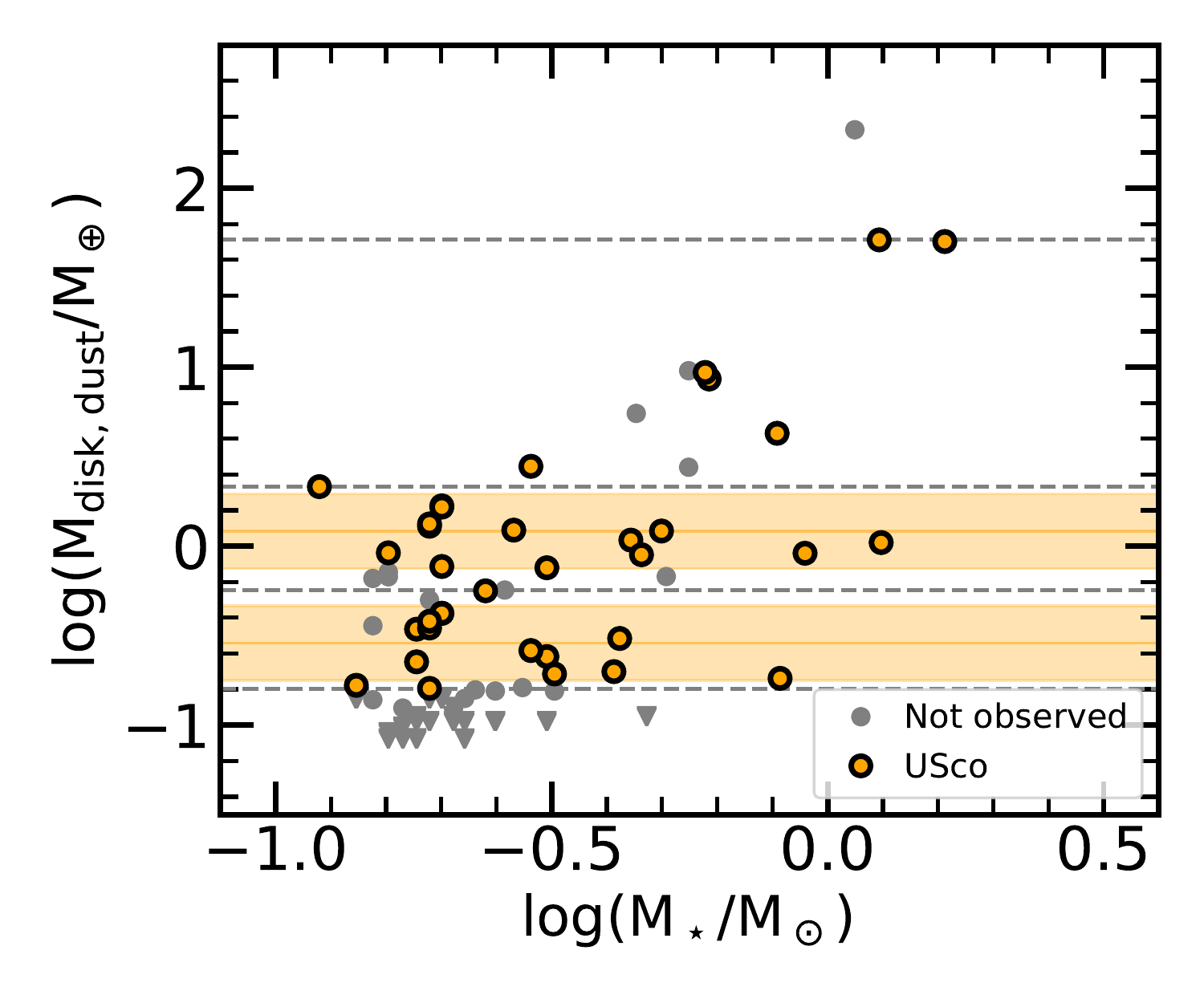}
\caption{Disk mass vs stellar mass after correcting the disk masses for the Gaia estimated distances and using the stellar masses derived here. Orange circles are used for targets observed with X-Shooter, gray symbols for targets not observed, circles for ALMA detections, downward facing triangles for ALMA upper limits. The shaded regions delimit the disk mass ranges where the sample is complete with respect to the \citet{barenfeld16} one. The dashed gray lines delimit the bins used in the discussion. For the objects not observed with X-Shooter, the distances, disk, and stellar masses from the literature are adopted.
     \label{fig::mdisk_mstar}}
\end{figure}

Our main goal is to quantify the median values and the spread of \macc \ in the \macc-\mdisk \ relation. For this reason, and given the allocated telescope time, we selected the stars with disks in two representative bins of \mdust \ for which we have an almost complete sample (Fig.~\ref{fig::mdisk_mstar}) compared to the \citet{barenfeld16} one. When we originally selected the sample, Gaia DR2 was not yet available. As a result of the revised distances, the completeness of our sample is not 100\% in the two disk mass bins 
0.16$\lesssim$\mdust/$M_\oplus\lesssim$0.563 and 0.75$\leq$\mdust/$M_\oplus\leq$1.957. 
On top of the targets in these two mass bins, we include in the analysis stars in Upper Scorpius that were observed in our previous observing run, as described in the next subsection. The disk mass of these additional targets is outside the boundaries of the two disk mass bins just introduced, and are mainly at higher disk masses.

Considering the samples from the two programs and the correction done using the information from Gaia, the completeness of our sample is as follows. Among the whole population of disks observed by \citet{barenfeld16} in Upper Scorpius with 
0.16$\lesssim$ \mdust/$M_\oplus$ $\lesssim$2.153, we obtained spectra for 28/36 of them. On top of that, we also observed 6/10 of the more massive disks. The sample includes two IR-classified transition disks (2MASS J16042165-2130284, 2MASS J16062196-1928445) and one additional transition disk resolved by ALMA (2MASS J15583692-2257153, \citealt{andrews18,ansdell20}), five ``evolved'' disks, meaning those with little infrared excess, and 26 full disks. Morever, one target that was not included in the sample of \citet{barenfeld16} was observed in our previous program. The latter is analyzed here, but cannot be included in the discussion due to the lack of a measured disk mass. Finally, one target (2MASSJ15354856-2958551) is a binary system that we resolved for the first time, and we associate the disk mass with both components. Therefore, the total number of targets for which we obtained the stellar and accretion properties here is 36, but the disk masses are available only for 35 of these. We verified that all the targets discussed here have parallaxes and proper motions compatible with being members of the Upper Scorpius association using the Gaia DR2 data.

\subsection{Observations}

All observations have been carried out with the X-Shooter spectrograph \citep{vernet11} on the VLT. 
Out of the 36 targets, eight were observed during our previous service mode program Pr.Id. 097.C-0378 (PI Manara), and 28 in the visitor mode program Pr.Id. 0101.C-0866 (PI Manara). In both programs, we obtained spectra both with narrow slits, to ensure a spectral resolution $R\gtrsim10000$ in the VIS and NIR arm ($\lambda > 500$nm), and $R\gtrsim5500$ in the UVB arm (300$\lesssim \lambda \lesssim$ 500 nm), as well as with 5.0\arcsec\ wide slits to correct the narrow slit spectra for slit losses. The slit was always oriented at parallactic angle, apart from the visual binary system where the slit was aligned to include both components. The log of the observations is discussed in App.~\ref{sect::logobs} and presented in Table~\ref{tab::log}.

\subsection{Data reduction}

Data reduction was carried out with the X-Shooter pipeline v2.9.3 \citep{xspipe} using the Reflex workflow v2.8.5 \citep{reflex}. The pipeline carries out the standard steps of flat, bias, and dark correction, wavelength calibration, spectral rectification and extraction of the 1D spectrum, and flux calibration using a standard star obtained in the same night. The 1D extraction of the spectra was carried out with IRAF from the rectified 2D spectrum in cases where the S/N of the UVB arm was low, and for resolved binaries.  Telluric correction was done using telluric standard stars observed close in time and airmass for the VIS arm, and molecfit \citep{molecfit1,molecfit2} for the NIR arms for both single stars and binaries. Finally, the spectra obtained with the narrow slits were rescaled to the wide slit ones to correct for slit losses. This procedure is the same as that used in previous works, for example, \citet{alcala17} and \citet{manara17a}.

\begin{figure}[]
\centering
\includegraphics[width=0.45\textwidth]{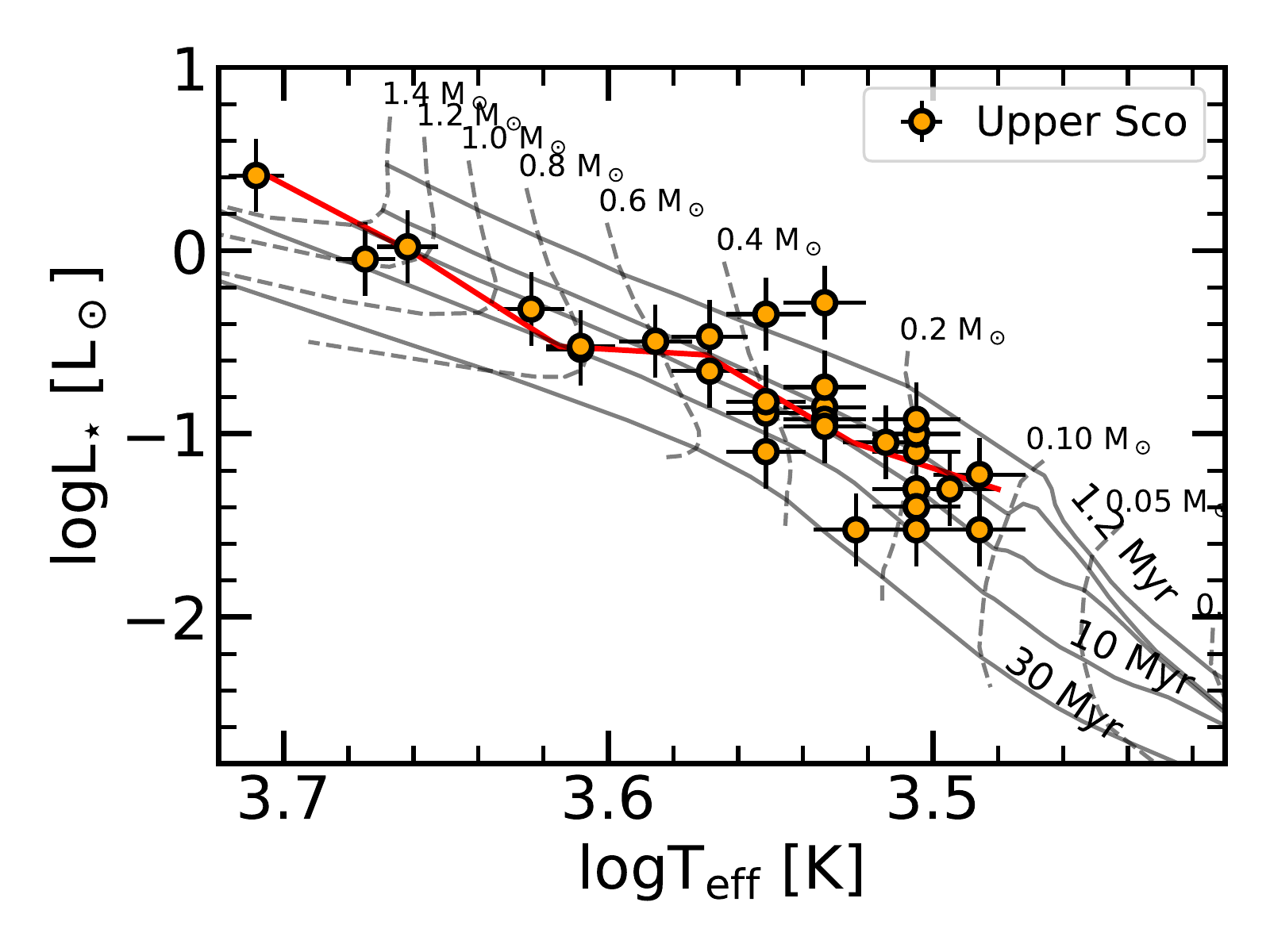}
\caption{HR diagram for objects in Upper Scorpius observed here. The evolutionary tracks are from \citet{B15}, with isochrones for 1.2, 3, 5, 10, and 30 Myr. The red line is the median of the \lstar \ in different $T_{\rm eff}$ bins. 
     \label{fig::HRD}}
\end{figure}


\section{Data analysis}\label{sect::analysis}

The analysis of the spectra to derive their stellar and accretion properties was carried out with the method described by \citet{manara13a}. In short, the observed spectrum is dereddened and fit with the sum of a photospheric template spectrum and a slab model to reproduce the continuum excess emission due to accretion. The grid of models used to find the best fit comprises various Class~III photospheric templates with different spectral types (SpT) from G- to late M-type taken from \citet{manara13a,manara17b}, different slab models, and extinction  values ($A_V$), assuming the reddening law by \citet{cardelli98} and $R_V=3.1$. The best fit of the Balmer continuum emission are shown in Figs.~\ref{fig::bj_1}-\ref{fig::bj_6}. The integrated flux of the best fit slab models gives an estimate of the excess luminosity due to accretion (\lacc), and the best fit normalization of the Class~III templates gives an estimate of the stellar luminosity (\lstar). By converting the SpT to $T_{\rm eff}$ using the relation by \citet{luhman03}, we are able to position the targets on the HR diagram (see Fig.~\ref{fig::HRD}) and obtain the stellar mass (\mstar) using the evolutionary models by \citet{B15} or \citet{siess00} (see Table~\ref{tab::res}). We note that our targets are located on the HRD typically between the 3 Myr and 10 Myr isochrones of the \citet{B15} models, with large spread at \mstar$\lesssim$0.4 \msun. The location of the targets on the HRD is thus in line with an age of $\sim$5-10 Myr for the region, and with an older age than other well-known star-forming regions, such as Lupus and Chamaeleon~I, which show typically higher values of \lstar \ at any $T_{\rm eff}$ for objects with disks. Finally, \macc \ was obtained from the relation \macc = 1.25 $\cdot$ \lacc $R_\star/(G M_\star)$.  All the stellar and accretion values are given in Table~\ref{tab::res}.

As several emission lines are present in the spectra, we measure their luminosity and convert them in \lacc \ using the relations by \citet{alcala17}. For the stronger accretors (\lacc$\gtrsim 10^{-4}$ \lsun), the values of \lacc \ obtained from the fit described above or from the emission line fluxes are similar within the uncertainties, as is usually found in accreting young stellar objects \citep[e.g.,][]{HH08,alcala14,alcala17}. 
For lower values of \lacc \ and for $\sim$20\% of the targets, instead, the accretion luminosity inferred from the line luminosities are systematically higher than those derived from the excess continuum luminosity, typically by a factor $\sim$5-10. This is in line with what was already observed by \citet{alcala14}; that the line emission is a higher fraction of the excess continuum emission for targets with low \lacc, with the total line emission being comparable to the continuum emission at \lacc$\lesssim 10^{-4}$ \lsun. 
We defer discussing this point to a future paper. In the following, we assume that \lacc \ is the one measured from the excess continuum emission. 
Here, it is sufficient to say that if we were to replace the continuum  excess luminosity with the sum of the continuum excess plus line emission the results discussed in the following would not be affected. We nevertheless note that \macc \ could be underestimated for the objects with the lowest accretion rates, which are typically below the chromospheric noise. 
\begin{figure}[]
\centering
\includegraphics[width=0.45\textwidth]{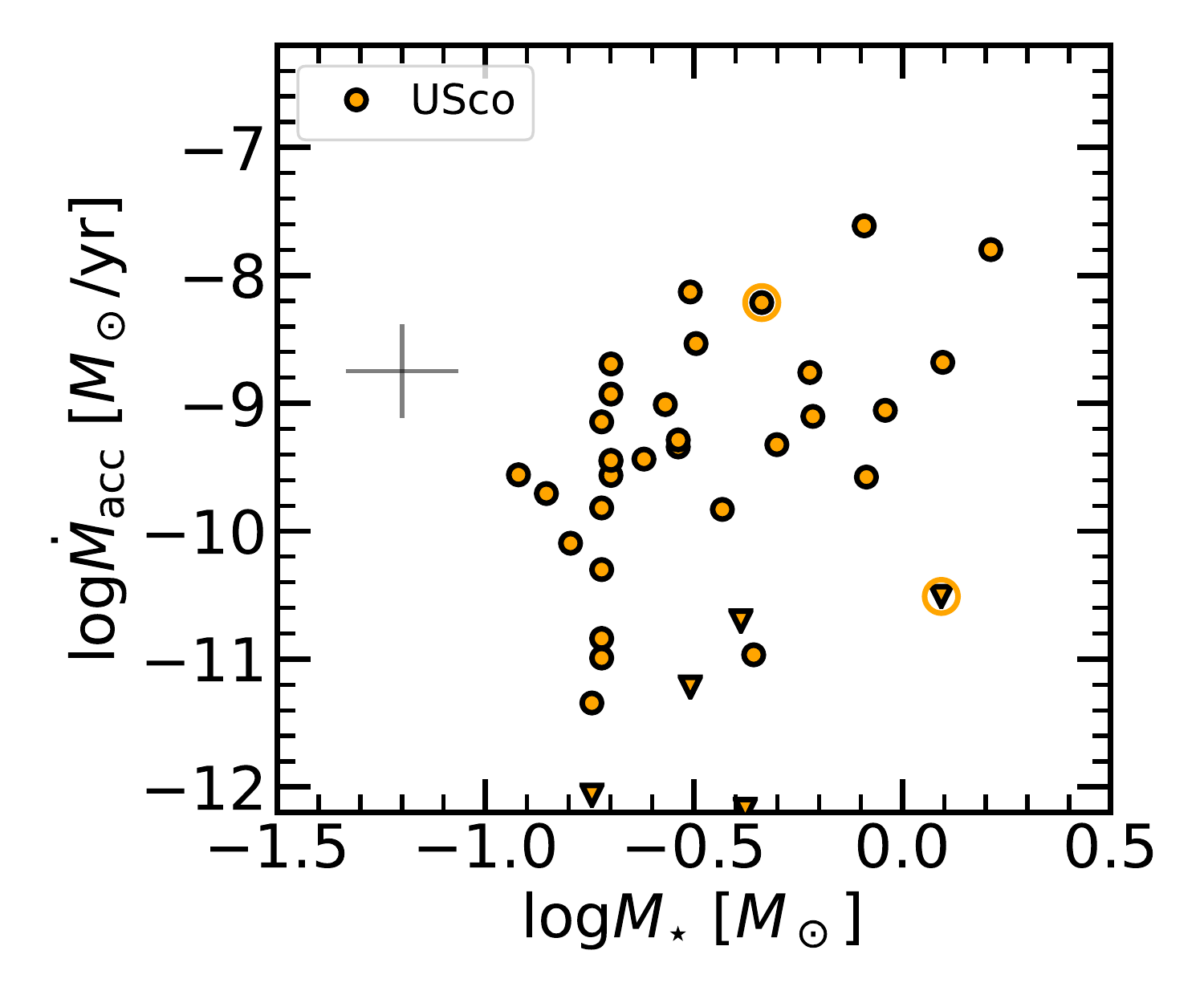}
\includegraphics[width=0.45\textwidth]{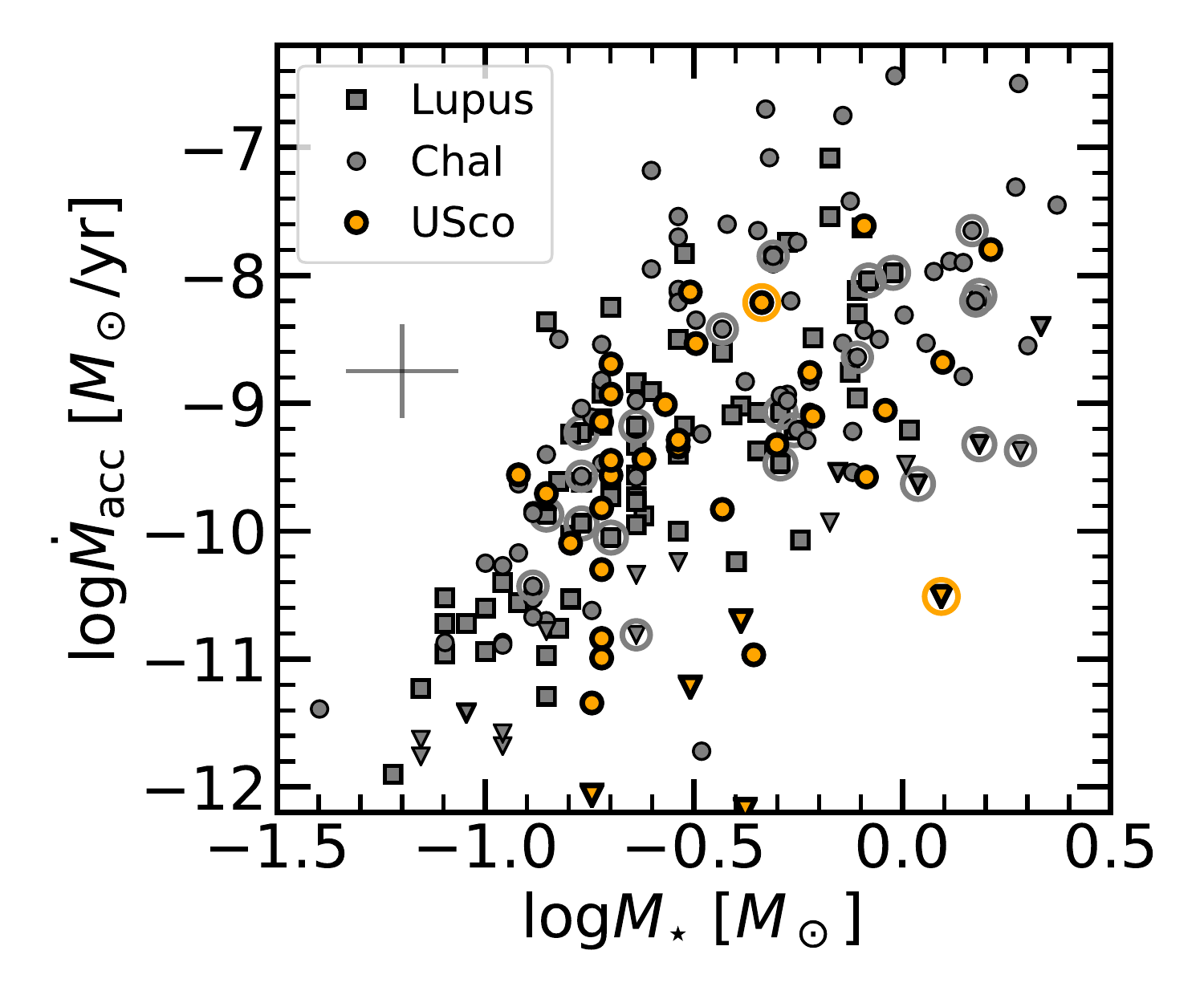}
\caption{Mass accretion rate vs stellar mass for the targets in the Upper Scorpius region (orange points, both in the upper and bottom panels) and for the targets in the Lupus and Chamaeleon~I regions (gray symbols, bottom panel). The downward facing triangles are used for non-accreting objects, transition disk objects are highlighted with a circle around their symbols. The cross indicates the typical errors on the measurements.
     \label{fig::macc_mstar}}
\end{figure}

For some of the targets with the lowest measured accretion rates, the ratio \lacc/\lstar \ falls below the typical values for chromospheric emission for their spectral type \citep{manara13a,manara17b}. In particular, five targets are significantly below this chromospheric emission noise when considering the continuum emission, and below or compatible with this noise when considering the line emission. We define these five targets as possible non-accretors (see Table~\ref{tab::res}), in line with previous work \citep[e.g.,][]{alcala14,alcala17,manara16a,manara17a}. The measured excess emission in these objects is considered in the analysis as an upper limit on the accretion rate, however, as discussed by \citet{manara17a}, the measured excess emission could be contaminated by other processes, in particular chromospheric emission. No excess in the Balmer continuum region with respect to a photosphere is detectable for these targets (see Figs.~\ref{fig::bj_1}-\ref{fig::bj_6}), in line with other estimates to confirm the accretion status of a young stellar object \citep[e.g.,][]{HH08,dealb20}. 
We note that for the non-accreting targets in the Upper Scorpius region, the measured upper limit on \lacc, and thus \macc, is generally lower by $\sim$0.5-1 dex than what is measured in similarly non-accreting targets in the Lupus and Chamaeleon~I regions. 
Since this estimate comes from the continuum emission fit and we noticed that the contribution of the lines is an higher fraction of the total excess emission at low \lacc{}, a small additional contribution from the line emission is also possible and would make these upper limits slightly higher, but still lower than those in younger regions.

The analysis of the spectra with the ROTFIT tool \citep{frasca17} leads to values of $T_{\rm eff}$ for the targets in line with those from the fitting procedure described before. The discussion of these results is deferred to a future work.

\begin{figure}[]
\centering
\includegraphics[width=0.45\textwidth]{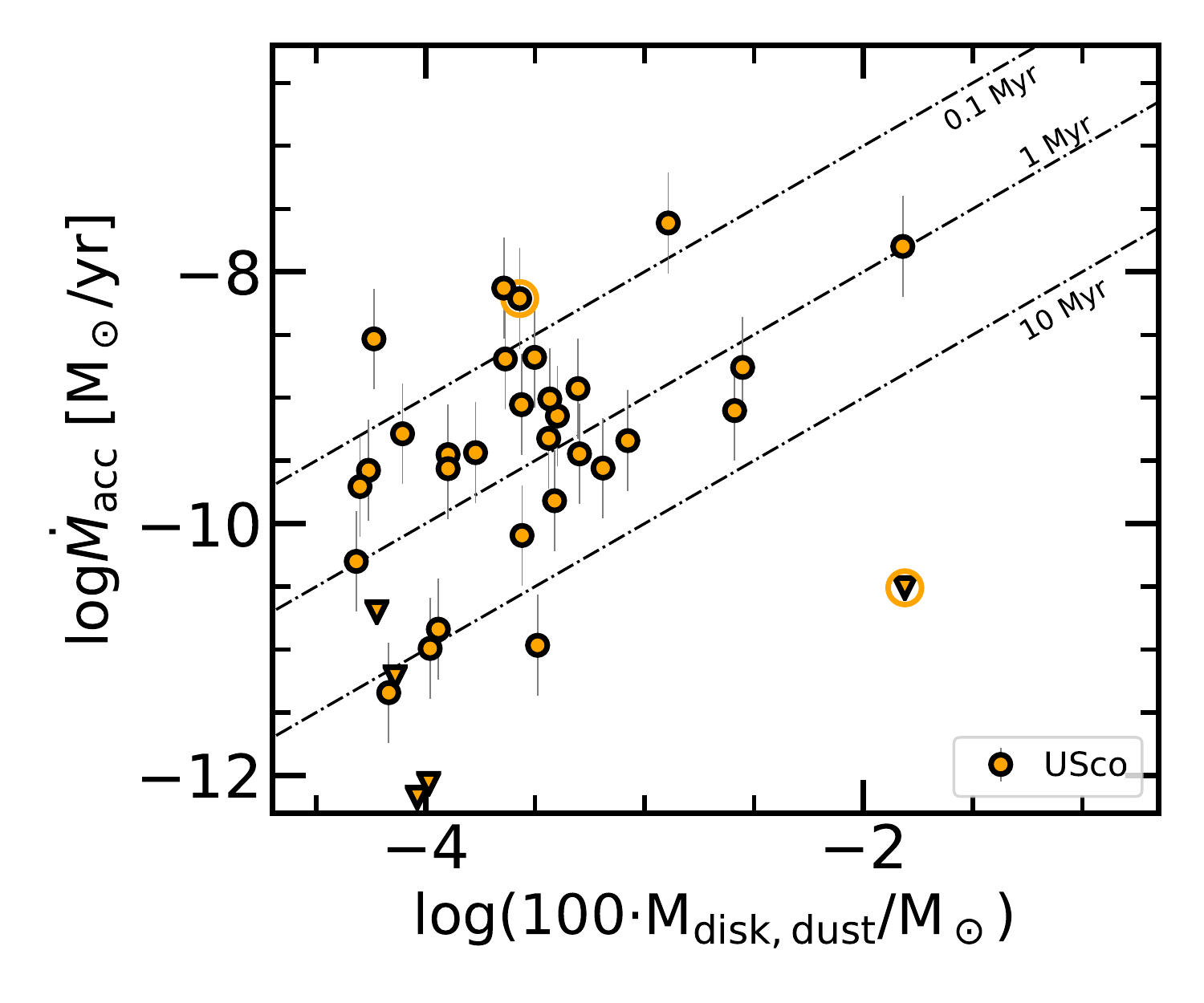}
\includegraphics[width=0.45\textwidth]{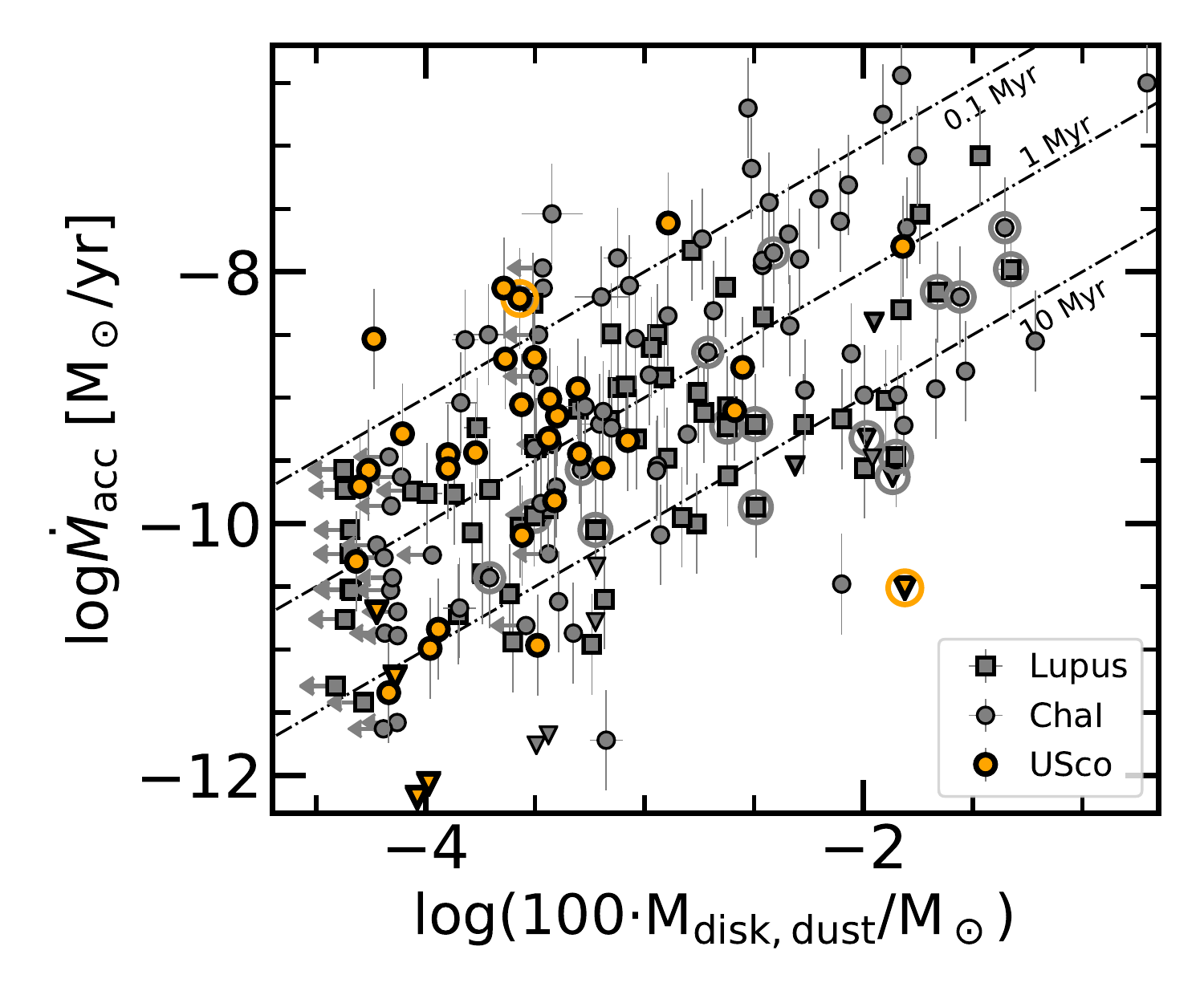}
\caption{Mass accretion rate vs disk mass for the targets in the Upper Scorpius region (orange points, both in the upper and bottom panels) and for the targets in the Lupus and Chamaeleon~I regions (gray symbols, bottom panel). The dot-dashed lines report different ratios of \mdisk/\macc: 0.1 Myr, 1 Myr, and 10 Myr, as labelled. Symbols as in Fig.~\ref{fig::macc_mstar}.
     \label{fig::macc_mdisk}}
\end{figure}


\section{Results}\label{sect::results}

Our analysis of the X-Shooter spectra of the targets allowed us to derive their stellar parameters, and, for the first time, their mass accretion rates. In this section, we discuss the relation between the following three parameters: the disk dust mass (\mdust), which is also a proxy of the total disk mass (\mdisk) assuming a constant gas-to-dust ratio of 100, the stellar mass (\mstar), and the mass accretion rate (\macc). 

\begin{figure*}[]
\centering
\includegraphics[width=\textwidth]{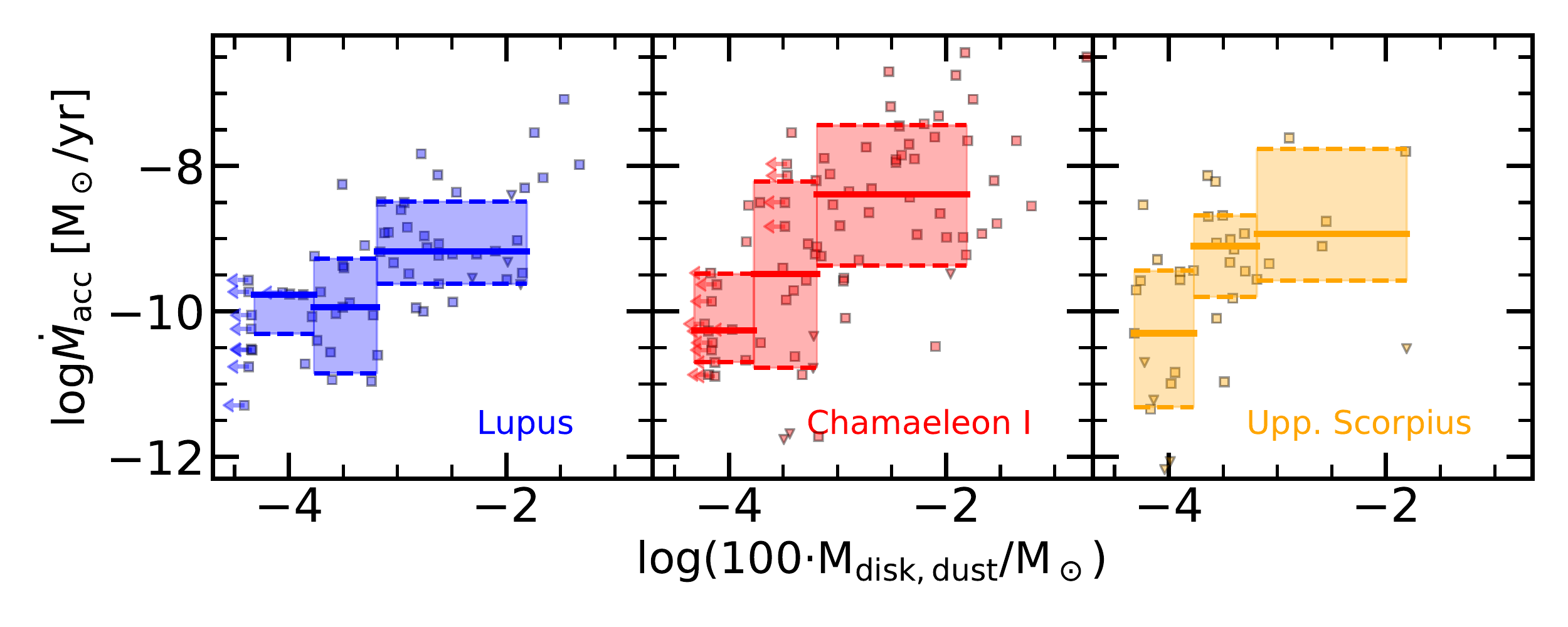}
\caption{Mass accretion rates vs disk dust masses for the targets in the Lupus and Chamaeleon~I star-forming regions, and in the Upper Scorpius region. The dashed lines report the 16$^{\rm th}$ and 84$^{\rm th}$ percentiles, and the solid line the median of the distributions. 
     \label{fig::mstar_mdisk_comb}}
\end{figure*}

The distribution of the measured \lacc \ as a function of \lstar \ (see App.~\ref{app::lacc_lstar}), as well as the one of \macc \ vs \mstar \ (Fig.~\ref{fig::macc_mstar}) reveals a great similarity with the relations observed in the younger Lupus and Chamaeleon~I star-forming regions \citep{alcala14,alcala17,manara16a,manara17a}. Indeed, the values of \macc \ measured in accreting objects with disks in the Upper Scorpius region show both similar values and a similar large ($\sim$1-2 dex) spread of \macc \ at any \mstar \ to the ones in these younger regions. One difference we note is that the maximum values of \macc \ measured in the Upper Scorpius sample (\macc$\sim 3\cdot 10^{-8} M_\odot$/yr) is in line with the maximum values measured in Lupus, but lower than the maximum values measured in Chamaeleon~I (\macc$\sim 3\cdot 10^{-7} M_\odot$/yr). However, as we discuss in the following, this might be an effect of the incompleteness of our sample at any given \mstar, as we only selected the targets based on their disk masses.

Similarly, the distribution of the data for the Upper Scorpius targets on the \macc-\mdisk \ plane (Fig.~\ref{fig::macc_mdisk}) is in overall agreement with the values measured in the younger star-forming regions of Lupus and Chamaeleon~I \citep{manara16b,mulders17}. A linear fit with the \emph{linmix} tool, which considers uncertainties on both axes and nondetections \citep{kelly07}, derives a similar slope ($0.8\pm0.4$) and spread ($\sigma=1.3$) on the Upper Scorpius sample  as the relation found in the younger regions \citep{manara16b,mulders17,manara19b}. However, the 
different level of completeness in the various bins of \mdisk can impact this result.

The samples in the younger Chamaeleon~I and Lupus star-forming regions include $>$90\% of the objects with IR-excess, such as a disk, in these regions \citep{alcala17,manara17a,pascucci16,ansdell16,ansdell18}. On the other hand, our sample in the Upper Scorpius region is, by construction, not similarly complete. Indeed, we selected only the most massive objects with IR-excess, making the sample complete at a similar level only in small ranges of \mdisk \ (see Sect.~\ref{sect::data}). 
In order to minimize the effects of incompleteness on the sample, we compared the median values and the spread of the \macc-\mdisk \ relation 
in the range of \mdisk,{} where the sample in the Upper Scorpius region is $\sim$80\% complete with respect to the initial sample of \citet{barenfeld16}. We thus selected the bins of \mdisk{} to carry out the analysis as reported in Table~\ref{tab::medians} and shown on Fig.~\ref{fig::mdisk_mstar} , such that three of these bins cover the \mdisk{} range with the highest completeness for the sample in the Upper Scorpius region.
In the first and second of the chosen bins, the sample in the Upper Scorpius region is $\sim$80\% complete; in the third one, the sample completeness is 60\%. 
These bins were then used to calculate the medians for the observed \macc, shown in  Fig.~\ref{fig::mstar_mdisk_comb}.

The comparison between the three datasets, presented in Fig.~\ref{fig::mstar_mdisk_comb} and reported also in Table~\ref{tab::medians}, shows that the median values of \macc \ are similar in the three regions, although typically slightly smaller for Lupus. The spread of \macc, measured as the difference between the 84$^{\rm th}$ and 16$^{\rm th}$ percentile of the distribution in any bin, is typically slightly larger in the Chamaeleon~I and Upper Scorpius regions ($\sim1.6-1.7$ dex) than in the Lupus region ($\sim1$ dex).

\begin{table*}  
\begin{center} 
\renewcommand{\arraystretch}{1.5}
\footnotesize 
\caption{\label{tab::medians} Median values for the \macc - \mdisk \ relation } 
\begin{tabular}{c|ccc|ccc|ccc | cc | cc } 
\hline \hline 
 Disk mass bin                  &       \multicolumn{3}{c}{Lupus} & \multicolumn{3}{c}{Chamaeleon~I} & \multicolumn{3}{c}{Upper Sco} & \multicolumn{2}{c}{Viscous 1 Myr} & \multicolumn{2}{c}{Viscous 8 Myr} \\
                        &       Median & Spr. & N$_{\rm data}$ & Median & Spr. & N$_{\rm data}$ & Median & Spr. & N$_{\rm data}$ & Median & Spr. & Median & Spr.  \\
\hline

$4.8\cdot 10^{-5} - 1.7\cdot 10^{-4}$   & $-$9.77 & 0.55 &5/0/1 & $-$10.26 & 1.22 &14/0/11 & $-$10.30 & 1.87 &15/4/0 & $-$10.01 & 1.11 &  $-$11.04 &  1.56 \\
$1.7\cdot 10^{-4} - 6.47\cdot 10^{-4}$ & $-$9.94 & 1.58 &15/0/0 & $-$9.48 & 2.56 &20/4/4 & $-$9.10 & 1.12  &14/0/0 & $-$9.53 & 0.76 &  $-$10.50 & 1.18 \\
$6.47\cdot 10^{-4} - 1.55\cdot 10^{-2}$ & $-$9.17 & 1.13 &32/4/0 & $-$8.39 & 1.93  &36/1/0 & $-$8.93 & 1.81 &6/1/0 & $-$8.91 & 1.35 & $-$9.65 & 1.08 \\

\hline 
\end{tabular} 
\tablefoot{\mdisk = 100 * \mdust in $M_\odot$. The table reports the values of log\macc, reported in $M_\odot$/yr, for the median, and for the spread of the distribution, defined as the difference between the 16$^{\rm th}$ and 84$^{\rm th}$ percentile of the distribution in a given bin of \mdisk. The latter is equivalent to a 2$\sigma$ spread. The N$_{\rm data}$ columns report the total number of tagets included in the bin / the number of non accretors in the bin / the number of undetected disks in the bin. } 
\end{center} 
\end{table*}  

\section{Discussion}\label{sect::discussion}

The results presented here show that the values of \macc \ measured in disk-hosting stars in a star-forming region with age$\sim$5-10 Myr are typically similar to those measured in disk-hosting stars in younger (age$<$3 Myr) regions. This is true both for the median values of \macc \ at given \mstar \ and/or \mdisk, and the spread of \macc \ values at given \mstar \ and/or \mdisk, which varies from one region to another but does not decrease with time. In particular, there are disks with high \macc$> 10^{-9} M_\odot$/yr and low disk masses, thus with \mdisk/\macc\ $\sim$0.1 Myr at all ages, even at $\sim5-10$ Myr. In the following, we discuss this result in light of some of the current models of disk evolution.

\subsection{The comparison with viscous evolution models}

\begin{figure*}[]
\centering
\includegraphics[width=0.7\textwidth]{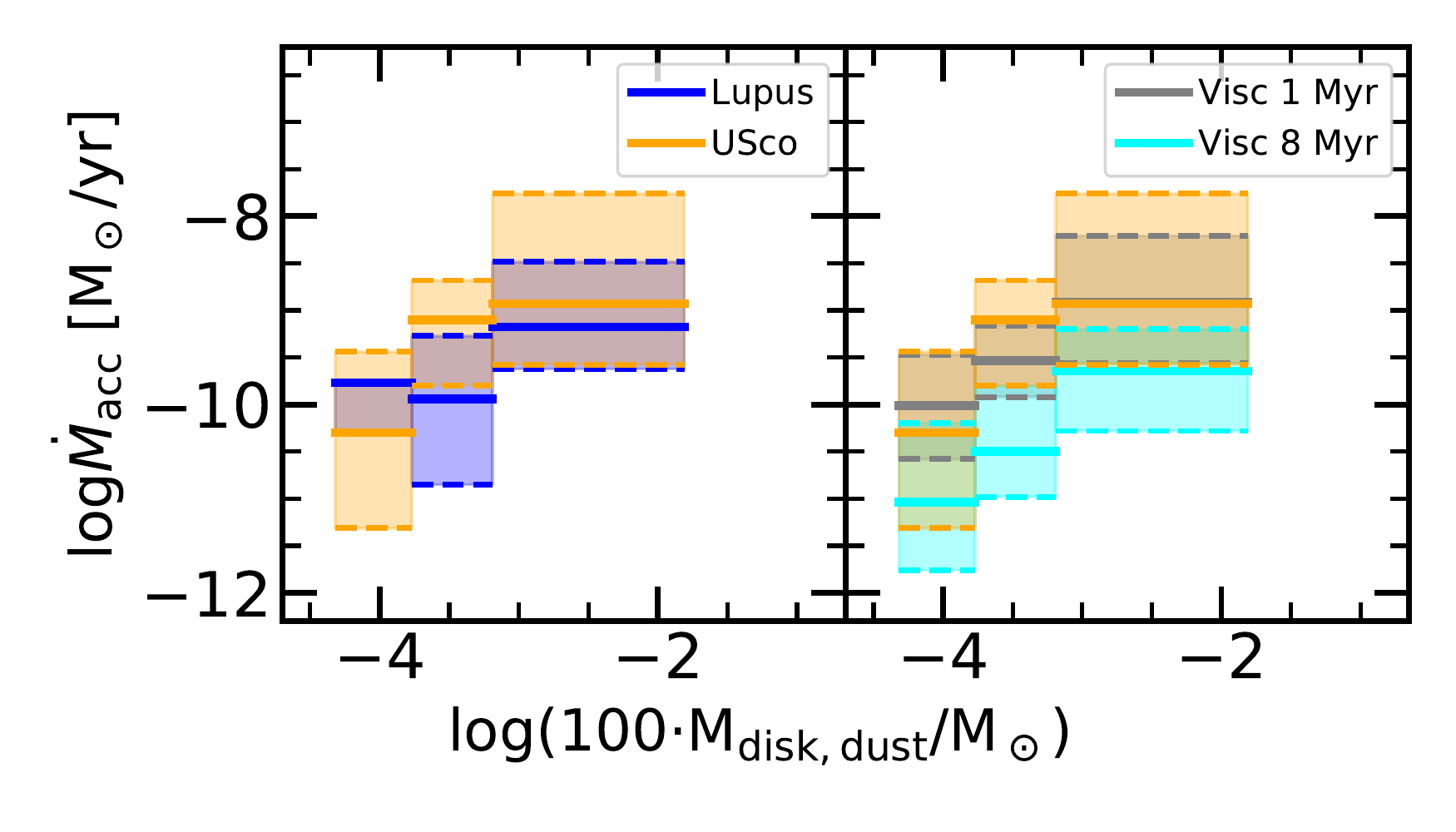}
\caption{Comparison between median and percentiles of the mass accretion rates as a function of disk mass for the Lupus and Upper Scorpius regions (left), and for the Upper Scorpius region and the expectations from viscous models at 1 Myr and 8 Myr (right).
     \label{fig::mstar_mdisk_comb_mod}}
\end{figure*}

The results shown here, taken at face value, are in contrast with a simple prescription of viscously evolving disks. A purely viscously evolving disk should have a value of \macc \ of the order of \mdisk \ divided by the age of the disk, as shown by \citet{jones12} and \citet{rosotti17}. This implies a tight correlation between these two quantities at ages much longer than the viscous timescale \citep[e.g.,][]{dullemond06,lodato17,mulders17}. In our data, both the values of \macc \ are higher in several targets than those expected given \mdisk \ in a viscous framework for disks of age $>$5 Myr, and the values of \macc \ are more spread than the tight correlation expected. These results are solid even when considering our selection biases, as we consider, in each \mdisk \ bin, a close to complete fraction of the known objects still retaining a disk -- traced by IR excess and ALMA detection.

To be able to reproduce the observed spread of the relation between \macc \ and \mdisk \ in the Lupus and Chamaeleon~I star-forming regions with viscous evolution models, both \citet{lodato17} and \citet{mulders17} needed to make several assumptions. First of all, the viscous timescale needed to be of the order of the age of the regions ($\sim$1 Myr). If this viscous timescale of $\sim$1 Myr were to be a universal value, this would imply that the correlation must be tight at ages$>5$ Myr. This is not observed here. Secondly, they needed to postulate a large dispersion of the model parameters; an age spread in the region, a distribution of initial conditions and of viscous timescales (or equivalently $\alpha$-viscosity parameter). When then the models were convolved with the observational uncertainties, both the observed slope and spread of the \macc-\mdisk \ relation were reproduced. \\
We tested our results against the best fitting viscous models for the Lupus star-forming regions obtained by \citet{lodato17}. These were described by a value of the exponent of the radial dependence of viscosity $\gamma$=1.5, a mean value of the viscosity timescale ($t_\nu$) of $\langle \log (t_\nu / {\rm yr})\rangle = 5.8$ with $\sigma_{t_\nu}$ = 1 dex, a mean age of the disks  $\langle \log (t / {\rm yr})\rangle = 5.9$ with $\sigma_t = 0.3$ dex, and further assuming $\langle \log (M_0 / M_\odot)\rangle = -2.2$, with $M_0$ being the initial disk mass of the models, and $\sigma_{M_0} = 0.2$ dex. We let these viscous models evolve in time until an age of 8 Myr. The expectations from these models are shown in Fig.~\ref{fig::mstar_mdisk_comb_mod} and reported in Table~\ref{tab::medians}. While the models predict a lower \macc \ at any \mdisk \ at 8 Myr compared to 1 Myr, the data show that the measured values of \macc \ in the Upper Scorpius region are closer to the expectations from models of 1 Myr old viscously evolving disks. In particular, the 84$^{th}$ percentile of the distribution of \macc \ expected by the models is always lower than the median value measured in the disks in the Upper Scorpius region. 
Also, as noted in Sect.~\ref{sect::results}, the spread of the values of \macc \ at any given \mdisk \ are similarly large in the older Upper Scorpius region as in the younger region of Chamaeleon~I, and larger than the viscous evolution model evolved at 8 Myr. This is particularly true when we compare the spread obtained fitting the model at 8 Myr using the {\em linmix} tool, $\sigma=0.4$ dex, with the data in the Upper Scorpius region, that have a spread with this method of $\sigma=1.3$ dex (see Sect.~\ref{sect::results}).
We thus observe that the models able to reproduce the observed \macc-\mdisk \ relation with pure viscous evolution for the Lupus region are not in agreement with the observations in the Upper Scorpius region, assuming only an age evolution from one region to another.

\subsection{The impact of photoevaporation and variable accretion}

The observations are even more discrepant from models predicting the \macc-\mdisk \ relation by means of both viscous evolution and internal photoevaporation. As shown by \citet{somigliana20}, the effect of photoevaporation is the reduction in the number of accreting targets at low disk masses and mass accretion rates to an extent that, by $\sim$10 Myr, only a fraction of massive disks still survive. This is not observed here, where we see low-mass disks with high \macc. It is unclear whether this disagreement is due to the fact that the models assume only one stellar mass and two fixed mass-loss rates, or whether this is an issue of internal photoevaporation models in general.

On the other hand, external photoevaporation would predict that the disks have low mass, while still low values of \mdisk/\macc$\sim$0.1 Myr \citep[see Fig.~\ref{fig::macc_mdisk}, and][]{rosotti17,sellek20}, more in line with what is observed here. In this context, it is worth mentioning that the environment of Upper Scorpius is different than the one of Chamaeleon~I and Lupus, having more nearby massive stars \citep[e.g.,][]{deZeeuw99}. In such an environment, the effect of external photoevaporation could have been relevant for the evolution of disks, possibly more than dynamical interactions \citep[e.g.,][]{winter18}. Whether this effect has been dominant for the evolution of the disks observed here is still an open question. Further modeling is mandatory here, but it is nevertheless puzzling how the mass accretion rates can be retained for such long time with so little disk mass available. 

A possible solution to the fact that the accretion rates measured here are high given the measured \mdisk\ might be variable accretion. However, studies in younger star-forming regions have shown that, in general, typical variation of \macc \ are $<$0.4 dex in most disks \citep[e.g.,][]{costigan14,venuti14}, with only a small fraction of targets showing extreme variability of \macc $>$1-2 dex \citep[e.g.,][]{audard14}. We could imagine that the variability is larger in the old region of Upper Scorpius, but in this case we could expect a larger dispersion in accretion rates than in younger regions, which, however, was not observed. Further studies on variability in older star-forming regions is needed. 
The mass-budget issue given the observed \macc \ and \mdisk \ is therefore important.
Indeed, under the assumption that \macc \ is constant with time, these high values of \macc \ would imply that over the lifetime of the disk $10^{-8} M_\odot$/yr $\cdot 10^7$ yr $\sim$ 0.1 \msun \ of disk gas mass is accreted from the disk onto the star. 
Assuming a gas-to-dust ratio of 100, this means that a total of \mdust$ \sim 10^{-3}$ \msun \ was accreted. This value is in line with the most massive disks observed in the Lupus and Chamaeleon~I disks, which could indeed be the progenitors of the survived disks observed here. Such high mass would probably imply that these disks were gravitationally unstable at the beginning of their lives \citep[e.g.,][]{KL16}. 
One possibility could be that accretion becomes active at later ages, as predicted by some models of MHD disk winds driven accretion \citep[e.g.,][for the case of a constant differential magnetic flux]{armitage13}.

\subsection{The need to account for dust evolution}

It is worth mentioning again that the assumption \mdisk = 100$\cdot$\mdust \ even after $\sim$5-10 Myr of disk evolution is possibly incorrect. As shown by global models of dust and gas evolution \citep[e.g.,][]{birnstiel2010,rosotti19}, dust radial drift is in general more efficient than gas accretion, implying that the dust-to-gas ratio is a decreasing function of time. Depending on the disk parameters (such as the efficiency of grain growth and the disk size), there can be an initial period of time, lasting $\sim$1-2 Myr, in which the assumption \mdisk$\sim$ 100$\cdot$\mdust \ is almost reasonable. However, this could not be the case for the Upper Scorpius targets, which are significantly older. 

At an age of $\sim$5-10 Myr, models tend to predict that the dust is depleted by a factor ranging from $\sim$10 to $\sim$100 \citep[e.g.,][]{birnstiel2010,rosotti19}. Such an increase in the gas-to-dust ratio to 1000 or more would make the median values of \mdisk/\macc \ more in line with expectations from viscous evolution by implying that the disks are substantially more massive than assumed here. A dedicated modeling effort would be needed to assess whether this is indeed a viable explanation, but this falls outside the scope of this paper. While this could reconcile the median values of \macc{} with viscous evolution models while making the disks in Upper Scorpius as massive as the younger disks in Chamaeleon~I and Lupus, it is unclear whether a better match with the observed age-independent spread could be obtained with such models. \citet{mulders17} already showed that a simple scatter in the values of the gas-to-dust ratio alone cannot reproduce the observed scatter in the \macc-\mdisk \ relation with no need for other sources of scatter, such as accretion variability. 

\subsection{Does the mass accretion rate decrease with time?}
The detection of strong accretors at later ages, and the connected hint of a lack of a general decrease of accretion rates with time when the targeted stars are still hosting a disk, has already been observed in different older star-forming regions: the nearby loose associations TWA \citep{venuti19} and $\eta$-Cha \citep{rugel18}, the more distant $\gamma$-Velorum cluster \citep{frasca15}, Orion OB1b and Orion OB1a associations \citep{ingleby14}, and even the very massive regions like NGC3603 or 30 Doradus \citep{demarchi17}. Individual targets have also been found to be still accreting at age$>$20 Myr \citep[e.g.,][]{mamajek02,zuckerman14,murphy18,lee20}.  
While this appears to be in contrast with evidence of a decrease of \macc \ with individual ages of young stars \citep[e.g.,][]{hartmann98,sicilia-aguilar10,antoniucci14,HCH16}, it should be noted that \citet{dario14} showed that correlated uncertainties on the determination of stellar parameters from the HR diagram can lead to spurious correlations between \macc \ and individual ages. Also, it is well known that the exact values of individual ages suffer from many uncertainties \citep[e.g.,][]{soderblom14}. The incompleteness of our sample does not allow us to draw final statements on this finding. 

We nevertheless would like to stress that we sampled only the older surviving disks. Indeed, it is well known that the fraction of disks and accretors decreases with time \citep[e.g.,][]{haisch01,hernandez07,fedele10}. Here, we can only consider the accretion rates of those disks that have survived until the age of the Upper Scorpius region, while the mass accretion rates of the other targets without IR- or mm-detected disk is by this age probably below the chromospheric activity limits of these old stars. While this is not an issue for the \macc-\mdisk \ relation, unless there is a population of massive disks with no accretion, it can impact measurements of typical \macc \ at different ages in populations of stars (see e.g., \citealt{sicilia-aguilar10}).

%

\section{Conclusions}\label{sect::conclusions}

We presented the analysis of the X-Shooter spectra of 36 young stellar objects with disks detected with ALMA in the $\sim$5-10 Myr old Upper Scorpius region. For the first time, the accretion rates for these targets were derived, together with their stellar properties. After re-scaling the values of the stellar, accretion, and disk properties with the new distances of the individual targets inferred from the Gaia DR2 parallaxes, we obtained the following results.
The dependence of \macc \ with \mstar \ and with \mdisk \ is similar in the Upper Scorpius region and in younger regions, such as Lupus and Chamaeleon~I. In particular, the median values of \macc \ at any given \mdisk \ are similar in the three regions, while the scatter of \macc \ varies from one region to another, but does not diminish with the age of the region. Both facts are in marked disagreement with simple predictions of viscous evolution models. The higher \macc \ values than predicted by viscous models for a given \mdisk{} could maybe be explained if the gas-to-dust ratio increases with time, as is expected by a radial-drift-dominated dust evolution process. 

The difficulties of simple disk viscous evolution models to explain our results stress the need to develop alternative models in more detail, such as those where the accretion through the disk is driven by MHD disk winds \citep[e.g.,][]{armitage13,bai13} coupled with global models of dust evolution, so that they could be validated against the existing body of observations.

On the observational side, future work should focus on completing the survey of \macc \ in older regions even in targets whose disks are not detected at millimetre wavelengths. At the same time, deep surveys of the gas emission in both young ($\sim$1-2 Myr) disks and disks with age $>$5 Myr are mandatory to establish whether the results presented here are due to a different process than viscous evolution, or to the outcome of the evolution of dust in disks.

\begin{acknowledgements}
We thank the anonymous referee for a detailed and constructive report that allowed us to improve the clarity of the paper. 
We thank S. Barenfeld and P. Cazzoletti for their help in the selection of the targets for the observations that were used for this work. 
CFM acknowledges an ESO fellowship.
This project has received funding from the European Union's Horizon 2020 research and innovation programme under the Marie Sklodowska-Curie grant agreement No 823823 (DUSTBUSTERS).
This work was partly supported by the Deutsche Forschungs-Gemeinschaft (DFG, German Research Foundation) - Ref no. FOR 2634/1 TE 1024/1-1. 
This work has made use of data from the European Space Agency (ESA) mission
{\it Gaia} (\url{https://www.cosmos.esa.int/gaia}), processed by the {\it Gaia}
Data Processing and Analysis Consortium (DPAC,
\url{https://www.cosmos.esa.int/web/gaia/dpac/consortium}). Funding for the DPAC
has been provided by national institutions, in particular the institutions
participating in the {\it Gaia} Multilateral Agreement. This work is part of the research programme VENI with project number 016.Veni.192.233, which is (partly) financed by the Dutch Research Council (NWO).
This work has been supported by the project PRIN-INAF Main Stream 2018 "Protoplanetary disks seen through the eyes of new generation instruments".

\end{acknowledgements}


\begin{table*}  
\begin{center} 
\footnotesize 
\caption{\label{tab::res} Stellar, disk, and accretion properties for the targets in the Upper Scorpius region } 
\begin{tabular}{l|c|c| ccc| cccc l c   } 
\hline \hline 
 Name                   &       disk &  dist      &     SpT  &  $T_{\rm eff}$ & $A_V$  &  \lstar & log\lacc   & \mstar &      \macc   & Acc?   &      \mdust \\
 &                      type &  [pc] &  & [K]  & [mag]  & \lsun  & \lsun  & \msun  & \msun/yr  &  & \msun \\
\hline
2MASSJ15534211-2049282          &  Full         & 136  $\pm$  4 &       M4    & 3270 & 1.2 &  0.09 & -2.6 &  0.24  &   3.66$\cdot10^{-10}$  &  Y   &                       1.69$\cdot10^{-06}$ \\
2MASSJ15583692-2257153          &  Full         & 166  $\pm$  4 &       K0    & 5110 & 0.0 &  2.57 & -0.5 &  1.63$^*$  &   1.59$\cdot10^{-08}$  &  Y &                       1.51$\cdot10^{-04}$ \\
2MASSJ16001844-2230114          &  Full         & 138  $\pm$  9 &       M4.5  & 3200 & 0.8 &  0.08 & -1.9 &  0.20  &   2.03$\cdot10^{-09}$  &  Y     &                       2.31$\cdot10^{-06}$ \\
2MASSJ16035767-2031055          &  Full         & 143  $\pm$  1 &       K6    & 4205 & 0.7 &  0.48 & -1.8 &  0.91  &   8.81$\cdot10^{-10}$  &  Y   &                       2.74$\cdot10^{-06}$ \\
2MASSJ16035793-1942108          &  Full         & 158  $\pm$  2 &       M2    & 3560 & 0.3 &  0.13 & -5.1 &  0.42  &   6.69$\cdot10^{-13}$  &  N   &                       9.16$\cdot10^{-07}$ \\
2MASSJ16041740-1942287          &  Full         & 161  $\pm$  2 &       M3    & 3415 & 0.7 &  0.14 & -4.3 &  0.31  &   6.04$\cdot10^{-12}$  &  N   &                       7.26$\cdot10^{-07}$ \\
2MASSJ16041893-2430392          &  \nodata       & 145   & M2   &  3560 & 0.3 &  0.45 & -3.1 &  0.37  &   1.48$\cdot10^{-10}$  &  Y       &                       \nodata    \\
2MASSJ16042165-2130284          &  Transitional & 150  $\pm$  1 &       K3    & 4730 & 1.4 &  0.90 & -3.2 &  1.24  &   3.09$\cdot10^{-11}$  &  N   &                       1.55$\cdot10^{-04}$ \\
2MASSJ15354856-2958551\_E  &  Full (binary)          & 145   & M4.5 &  3200 & 0.0 &  0.10 & -2.8 &  0.20  &   3.53$\cdot10^{-10}$  &  Y     &                       1.27$\cdot10^{-06}$ \\
2MASSJ15354856-2958551\_W  & \nodata~(binary)           & 145   & M4.5 &  3200 & 0.0 &  0.10 & -2.9 &  0.20  &   2.73$\cdot10^{-10}$  &  Y       &                       1.27$\cdot10^{-06}$ \\
2MASSJ15514032-2146103     &  Evolved       & 142  $\pm$  2 &   M4.5  & 3200 & 0.3 &  0.05 & -3.5 &  0.19  &   5.01$\cdot10^{-11}$  &  Y     &                       4.82$\cdot10^{-07}$ \\
2MASSJ15530132-2114135     &  Full          & 146  $\pm$  2 &   M4.5  & 3200 & 0.8 &  0.05 & -3.0 &  0.19  &   1.52$\cdot10^{-10}$  &  Y     &                       3.88$\cdot10^{-06}$ \\
2MASSJ15582981-2310077     &  Full          & 147  $\pm$  3 &   M4.5  & 3200 & 1.0 &  0.05 & -2.3 &  0.19  &   7.16$\cdot10^{-10}$  &  Y     &                       4.00$\cdot10^{-06}$ \\
2MASSJ16014086-2258103     &  Full          & 145   & M3   &  3415 & 1.2 &  0.12 & -1.2 &  0.31  &   7.42$\cdot10^{-09}$  &  Y   &                       2.28$\cdot10^{-06}$ \\
2MASSJ16020757-2257467     &  Full          & 140  $\pm$  1 &   M2    & 3560 & 0.4 &  0.08 & -3.8 &  0.44  &   1.08$\cdot10^{-11}$  &  Y     &                       3.25$\cdot10^{-06}$ \\
2MASSJ16024152-2138245     &  Full          & 142  $\pm$  2 &   M5.5  & 3060 & 0.6 &  0.03 & -2.9 &  0.12  &   2.76$\cdot10^{-10}$  &  Y     &                       6.46$\cdot10^{-06}$ \\
2MASSJ16054540-2023088     &  Full          & 145  $\pm$  2 &   M4.5  & 3200 & 0.6 &  0.10 & -2.8 &  0.20  &   3.58$\cdot10^{-10}$  &  Y     &                       5.05$\cdot10^{-06}$ \\
2MASSJ16062196-1928445     &  Transitional  & 145   & M1   &  3705 & 0.8 &  0.34 & -1.3 &  0.46  &   6.13$\cdot10^{-09}$  &  Y   &                       2.69$\cdot10^{-06}$ \\
2MASSJ16063539-2516510     &  Evolved       & 139  $\pm$  3 &   M4.5  & 3200 & 0.0 &  0.03 & -5.1 &  0.18  &   8.62$\cdot10^{-13}$  &  N     &                       1.03$\cdot10^{-06}$ \\
2MASSJ16064385-1908056     &  Evolved       & 144  $\pm$  7 &   K7    & 4060 & 0.4 &  0.29 & -2.3 &  0.82  &   2.65$\cdot10^{-10}$  &  Y     &                       5.48$\cdot10^{-07}$ \\
2MASSJ16072625-2432079     &  Full          & 143  $\pm$  2 &   M3    & 3415 & 0.7 &  0.18 & -2.6 &  0.29  &   4.56$\cdot10^{-10}$  &  Y     &                       8.39$\cdot10^{-06}$ \\
2MASSJ16081566-2222199     &  Full          & 140  $\pm$  2 &   M2    & 3560 & 0.5 &  0.15 & -3.7 &  0.41  &   1.99$\cdot10^{-11}$  &  N     &                       5.98$\cdot10^{-07}$ \\
2MASSJ16082324-1930009     &  Full          & 138  $\pm$  1 &   M0    & 3850 & 1.1 &  0.32 & -2.0 &  0.61  &   7.90$\cdot10^{-10}$  &  Y     &                       2.58$\cdot10^{-05}$ \\
2MASSJ16082751-1949047     &  Evolved       & 145   & M5.5 &  3060 & 0.6 &  0.06 & -3.1 &  0.14  &   1.97$\cdot10^{-10}$  &  Y   &                       5.01$\cdot10^{-07}$ \\
2MASSJ16090002-1908368     &  Full          & 139  $\pm$  3 &   M4.5  & 3200 & 0.3 &  0.05 & -4.2 &  0.19  &   1.02$\cdot10^{-11}$  &  Y     &                       1.05$\cdot10^{-06}$ \\
2MASSJ16090075-1908526     &  Full          & 138  $\pm$  1 &   M0    & 3850 & 1.0 &  0.32 & -1.7 &  0.60  &   1.74$\cdot10^{-09}$  &  Y     &                       2.81$\cdot10^{-05}$ \\
2MASSJ16095361-1754474     &  Full          & 158  $\pm$  5 &   M4.5  & 3200 & 0.5 &  0.04 & -4.5 &  0.18  &   4.54$\cdot10^{-12}$  &  Y     &                       6.78$\cdot10^{-07}$ \\
2MASSJ16104636-1840598     &  Full          & 143  $\pm$  3 &   M4.5  & 3200 & 1.2 &  0.04 & -3.9 &  0.19  &   1.45$\cdot10^{-11}$  &  Y     &                       1.14$\cdot10^{-06}$ \\
2MASSJ16111330-2019029     &  Full          & 155  $\pm$  1 &   M3.5  & 3340 & 0.6 &  0.03 & -1.9 &  0.27  &   9.77$\cdot10^{-10}$  &  Y     &                       3.69$\cdot10^{-06}$ \\
2MASSJ16123916-1859284     &  Full          & 139  $\pm$  2 &   M1    & 3705 & 0.6 &  0.22 & -2.3 &  0.50  &   4.75$\cdot10^{-10}$  &  Y     &                       3.65$\cdot10^{-06}$ \\
2MASSJ16133650-2503473     &  Full          & 145   & M3   &  3415 & 1.0 &  0.11 & -1.6 &  0.32  &   2.93$\cdot10^{-09}$  &  Y   &                       5.80$\cdot10^{-07}$ \\
2MASSJ16135434-2320342     &  Full          & 145   & M4.5 &  3200 & 0.3 &  0.12 & -2.3 &  0.20  &   1.18$\cdot10^{-09}$  &  Y   &                       4.97$\cdot10^{-06}$ \\
2MASSJ16141107-2305362     &  Full          & 145   & K4   &  4590 & 0.3 &  1.05 & -1.4 &  1.25  &   2.09$\cdot10^{-09}$  &  Y   &                       3.15$\cdot10^{-06}$ \\
2MASSJ16143367-1900133     &  Full          & 142  $\pm$  2 &   M3    & 3415 & 1.9 &  0.52 & -2.7 &  0.29  &   5.17$\cdot10^{-10}$  &  Y     &                       7.84$\cdot10^{-07}$ \\
2MASSJ16154416-1921171     &  Full          & 132  $\pm$  2 &   K7    & 4060 & 2.8 &  0.30 & -0.3 &  0.81  &   2.44$\cdot10^{-08}$  &  Y     &                       1.28$\cdot10^{-05}$ \\
2MASSJ16181904-2028479     &  Evolved       & 138  $\pm$  2 &   M5    & 3125 & 1.6 &  0.05 & -3.4 &  0.16  &   8.05$\cdot10^{-11}$  &  Y     &                       2.76$\cdot10^{-06}$ \\
\hline 
\end{tabular} 
\tablefoot{Disk type from \citet{barenfeld16,luhman12,carpenter06}.
Stellar properties obtained using the \citet{B15} evolutionary models, apart from the target 2MASSJ15583692-2257153, for which \citet{siess00} models were used since the stellar mass was higher than the maximum one modeled by \citet{B15}. 
Disk masses are updated from \citet{barenfeld16} using the distance inferred from the Gaia DR2 \citep{gaiadr2} parallaxes. When no uncertainties on the distance is reported, the mean distance to the targets of 145 pc was adopted. Possible non-accretors are reported with "N" in the Acc? column. The values reported here for the accretion rate of non-accretors are considered in this work as upper limits. } 
\end{center} 
\end{table*}  

\appendix

\section{Log of the observations}\label{sect::logobs}

The observations were carried out in two different observing programs. Eight targets were observed in the Service Mode program Pr.Id. 097.C-0378 (PI Manara) in the period from July to August 2016. Typically, these observations were carried out with image quality in the VIS arm of $\sim$1\arcsec~(see Table~\ref{tab::log}). The standard star observed at the beginning of the night as part of the standard calibration plan for X-Shooter was used for the flux calibration of the spectra. 

The remaining 28 targets discussed here were observed during the Visitor Mode program Pr.Id. 0101.C-0866 (PI Manara) carried out during the nights of May 19 and 20, 2018. Both nights had very good seeing, typically $<$0.5\arcsec, leading to image qualities in the VIS arm better than 1\arcsec \ in all cases but two. Small clouds (THN conditions) were present at the beginning of the nights, otherwise the nights were clear. We observed flux standard stars at the beginning and at the end of the night during the first night, and at the beginning, in the middle, and at the end of the night during the second night. The reduction led to consistent results with all standard stars. For the reduction of the data obtained in the first night, we adopted the standard star observed at the end of the night. For the data obtained in the first part of the second night we used the standard star observed in the middle of the night, while we used the one observed at the end of the night for the spectra obtained in the second half of the night, starting from and including 2MASS J16072625-2432079.

\begin{table*}  
\begin{center}  
\footnotesize  
\caption{\label{tab::log} Night log of the observations }  
\begin{tabular}{l|c|c| ccc | c }    
\hline \hline  
2MASS               &        Date of observation [UT]    & Exp. Time  & \multicolumn{3}{c}{Slit width [\arcsec]} &      I.Q. \\
& & [Nexp x (s)] & UVB & VIS & NIR & [\arcsec] \\
\hline 
J15534211-2049282       &       2016-07-24T04:02:25.209 &  4x450                 &       1.0  &          0.4   &         0.4   &         1.07      \\                     
J15583692-2257153       &       2016-07-25T03:37:53.376 &  4x120                 &       0.5  &          0.4   &         0.4   &         1.15      \\                     
J16001844-2230114       &       2016-08-09T01:56:46.720 &  4x450                 &       1.0  &          0.4   &         0.4   &         0.91      \\                     
J16035767-2031055       &       2016-08-18T02:04:50.750 &  4x150                 &       0.5  &          0.4   &         0.4   &         1.03      \\                     
J16035793-1942108       &       2016-08-09T00:06:57.436 &  4x450                 &       1.0  &          0.4   &         0.4   &         0.72      \\                     
J16041740-1942287       &       2016-08-07T23:57:24.090 &  4x450                 &       1.0  &          0.4   &         0.4   &         1.02      \\                     
J16041893-2430392       &       2016-08-26T02:20:38.193 &  4x150                 &       0.5  &          0.4   &         0.4   &         1.02      \\                     
J16042165-2130284       &       2016-08-18T02:43:53.101 &  4x150                 &       0.5  &          0.4   &         0.4   &         0.9       \\                     
J15354856-2958551\_E    &       2018-05-19T23:39:04.373 &  4x300                 &       1.0  &          0.4   &         0.4   &         0.42      \\                     
J15354856-2958551\_W    &       2018-05-19T23:39:04.373 &  4x300                 &       1.0  &          0.4   &         0.4   &         0.42      \\                     
J15514032-2146103       &       2018-05-21T02:38:56.215 &  4x675                 &       1.0  &          0.9   &         0.9   &         0.75      \\                     
J15530132-2114135       &       2018-05-20T02:06:26.327 &  4x675                 &       1.0  &          0.9   &         0.9   &         0.9       \\                     
J15582981-2310077       &       2018-05-20T03:07:24.550 &  4x630                 &       1.0  &          0.9   &         0.9   &         0.86      \\                     
J16014086-2258103       &       2018-05-20T08:38:14.117 &  4x150                 &       1.0  &          0.4   &         0.4   &         0.97      \\                     
J16020757-2257467       &       2018-05-20T00:39:53.156 &  4x195                 &       1.0  &          0.4   &         0.4   &         0.93      \\                     
J16024152-2138245       &       2018-05-21T05:22:25.139 &  4x675                 &       1.0  &          0.9   &         0.9   &         0.77      \\                     
J16054540-2023088       &       2018-05-20T08:04:28.437 &  4x300                 &       1.0  &          0.4   &         0.4   &         1.03      \\                     
J16062196-1928445       &       2018-05-20T00:15:54.949 &  4x120                 &       0.5  &          0.4   &         0.4   &         0.9       \\                     
J16063539-2516510       &       2018-05-20T07:10:36.261 &  4x525                 &       1.0  &          0.9   &         0.9   &         0.73      \\                     
J16064385-1908056       &       2018-05-21T00:37:34.606 &  4x120                 &       1.0  &          0.4   &         0.4   &         0.84      \\                     
J16072625-2432079       &       2018-05-21T06:51:13.403 &  4x225                 &       1.0  &          0.4   &         0.4   &         0.64      \\                     
J16081566-2222199       &       2018-05-21T01:11:07.195 &  4x195                 &       1.0  &          0.4   &         0.4   &         0.96      \\                     
J16082324-1930009       &       2018-05-21T07:19:48.545 &  4x120                 &       1.0  &          0.4   &         0.4   &         0.7       \\                     
J16082751-1949047       &       2018-05-21T01:39:00.144 &  4x450                 &       1.0  &          0.4   &         0.4   &         1.        \\                     
J16090002-1908368       &       2018-05-21T03:39:56.012 &  4x600                 &       1.0  &          0.9   &         0.9   &         0.7       \\                     
J16090075-1908526       &       2018-05-20T09:34:05.964 &  4x75                  &       1.0  &          0.4   &         0.4   &         0.9       \\                     
J16095361-1754474       &       2018-05-20T05:17:57.934 &  4x600                 &       1.0  &          0.9   &         0.9   &         0.46      \\                     
J16104636-1840598       &       2018-05-20T06:14:36.404 &  4x600                 &       1.0  &          0.9   &         0.9   &         0.74      \\                     
J16111330-2019029       &       2018-05-20T01:07:08.211 &  4x195                 &       1.0  &          0.4   &         0.4   &         0.99      \\                     
J16123916-1859284       &       2018-05-20T01:34:34.877 &  4x120                 &       1.0  &          0.4   &         0.4   &         1.09      \\                     
J16133650-2503473       &       2018-05-21T06:22:13.098 &  4x225                 &       1.0  &          0.4   &         0.4   &         0.63      \\                     
J16135434-2320342       &       2018-05-20T09:02:41.397 &  4x150                 &       1.0  &          0.4   &         0.4   &         0.9       \\                     
J16141107-2305362       &       2018-05-21T00:21:01.107 &  4x140                 &       0.5  &          0.4   &         0.4   &         1.38      \\                     
J16143367-1900133       &       2018-05-21T04:45:56.354 &  4x300                 &       1.0  &          0.4   &         0.4   &         0.62      \\                     
J16154416-1921171       &       2018-05-21T07:41:25.241 &  4x120                 &       1.0  &          0.4   &         0.4   &         0.71      \\                     
J16181904-2028479       &       2018-05-20T04:17:13.271 &  4x675                 &       1.0  &          0.9   &         0.9   &         0.82      \\                     
\hline 
\end{tabular} 
\tablefoot{Typical resolutions in the UVB arm are $R\sim$9700 for 0.5\arcsec \ wide slits, $R\sim$5400 for 1.0\arcsec \ wide slits; in the VIS arm $R\sim$18400 for 0.4\arcsec \ wide slits, $R\sim$8900 for 0.9\arcsec \ slits; in the NIR arm $R\sim$11600 for 0.4\arcsec \ wide slits, and $R\sim$5600 for 0.9\arcsec \ wide slits. I.Q. is the airmass corrected seeing.}
\end{center} 
\end{table*}

\subsection{Resolved binaries}
During the visitor mode observations, we resolved two close-by stars: 2MASS J15354856-2958551 and 2MASS J16054540-2023088. The former system is composed of two stars at about $\sim$1\arcsec~distance from each other in the west-east direction. They were observed by orienting the slit at position angle $-$105.27\degree, while the parallactic angle was $-$104.85\degree. The latter system is composed of two objects at 2.14\arcsec~distance, and they were both included in the slit oriented at position angle 54.4\degree. 
The two traces are resolved in both observations when using the narrow slits, while only for 2MASS J16054540-2023088 when using the wide slit. We manually extracted the two spectra from the pipeline reduced 2D spectra using IRAF\footnote{IRAF is distributed by the National Optical Astronomy Observatory, which is operated by the Association of Universities for Research in Astronomy (AURA) under a cooperative agreement with the National Science Foundation.}. 

In the case of 2MASS J15354856-2958551, both spectra are those of a young stellar object, showing  clear Lithium absorption lines and strong emission lines. We flux-calibrated the narrow slit spectra calculating the ratio between the combined ones in the large slit exposure and the sum of the separated spectra in the narrow slit exposure. 

On the other hand, in the case of J16054540-2023088, one spectrum is the one of a young stellar object, while the other one is an early-type background object. Indeed, the latter becomes fainter in the optical and near-infrared than the YSO. The location of the YSO is at the correct 2MASS coordinates. We thus flux calibrate the YSO spectrum taken with the narrow slit using the one with the wide slit for this object alone. 

\section{Information from Gaia}
We searched for the Gaia \citep{gaia} counterpart for our targets in the Gaia DR2 catalog \citep{gaiadr2}. 
Only six of our targets have no astrometric solutions, and in one case no matching with Gaia is found. In another case, the parallax is negative and the proper motion very different with respect to other objects in our sample (2MASS J16141107-2305362). In two cases, the matching is with separation $>$0.7\arcsec : one is for a component of the binary system, but in this case there is no astrometric solution, in the other case it is for 2MASS J16014086-2258103, and we do not consider the parallax value reliable.

We checked whether any of the targets shared similar astrometric parameters (proper motion, parallax, coordinates) with young stars in the $\rho$-Ophiucus region. While the proper motions are similar, the spatial location is different, and we consider all our targets bona-fide members of the Upper Scorpius association.

The distances to individual targets are reported in Table~\ref{tab::res} and are obtained by inverting the parallaxes. When no Gaia parallax is available we assumed $d$=145 pc.

\section{Additional plots}

\subsection{Best fit of the Balmer continuum emission}\label{app::BJ_plot}

In the following, we show the best fit of the Balmer continuum emission for the targets analyzed here. 

\begin{figure*}[]
\centering
\includegraphics[width=0.9\textwidth,page=1]{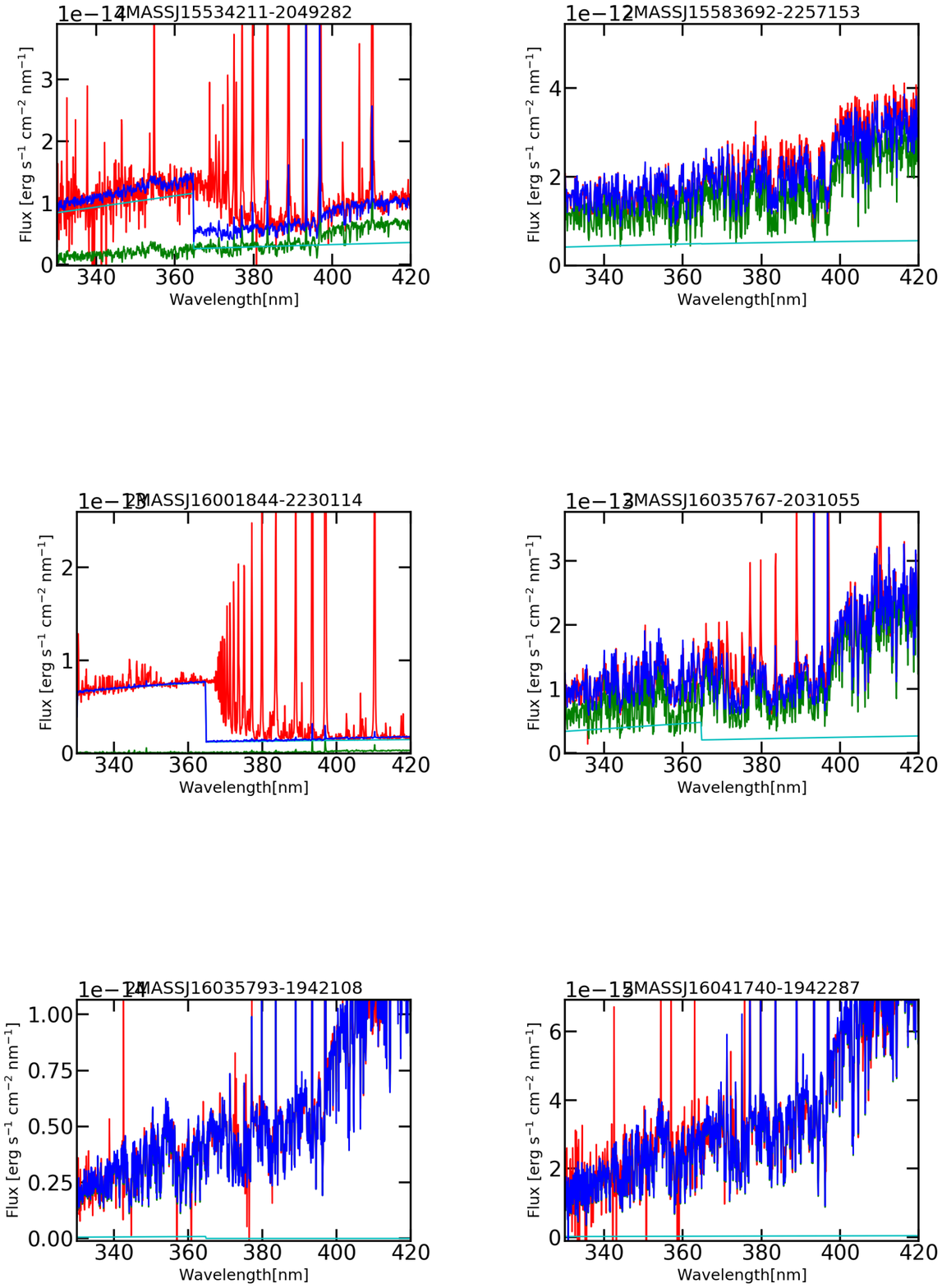}
\caption{Best fit of the continuum emission of the spectrum of the targets is shown in blue together with the dereddened spectrum of the target (red). The best fit is obtained by adding the best fit photospheric template (green) and the slab model (cyan).  
     \label{fig::bj_1}}
\end{figure*}

\begin{figure*}[]
\centering
\includegraphics[width=0.9\textwidth,page=2]{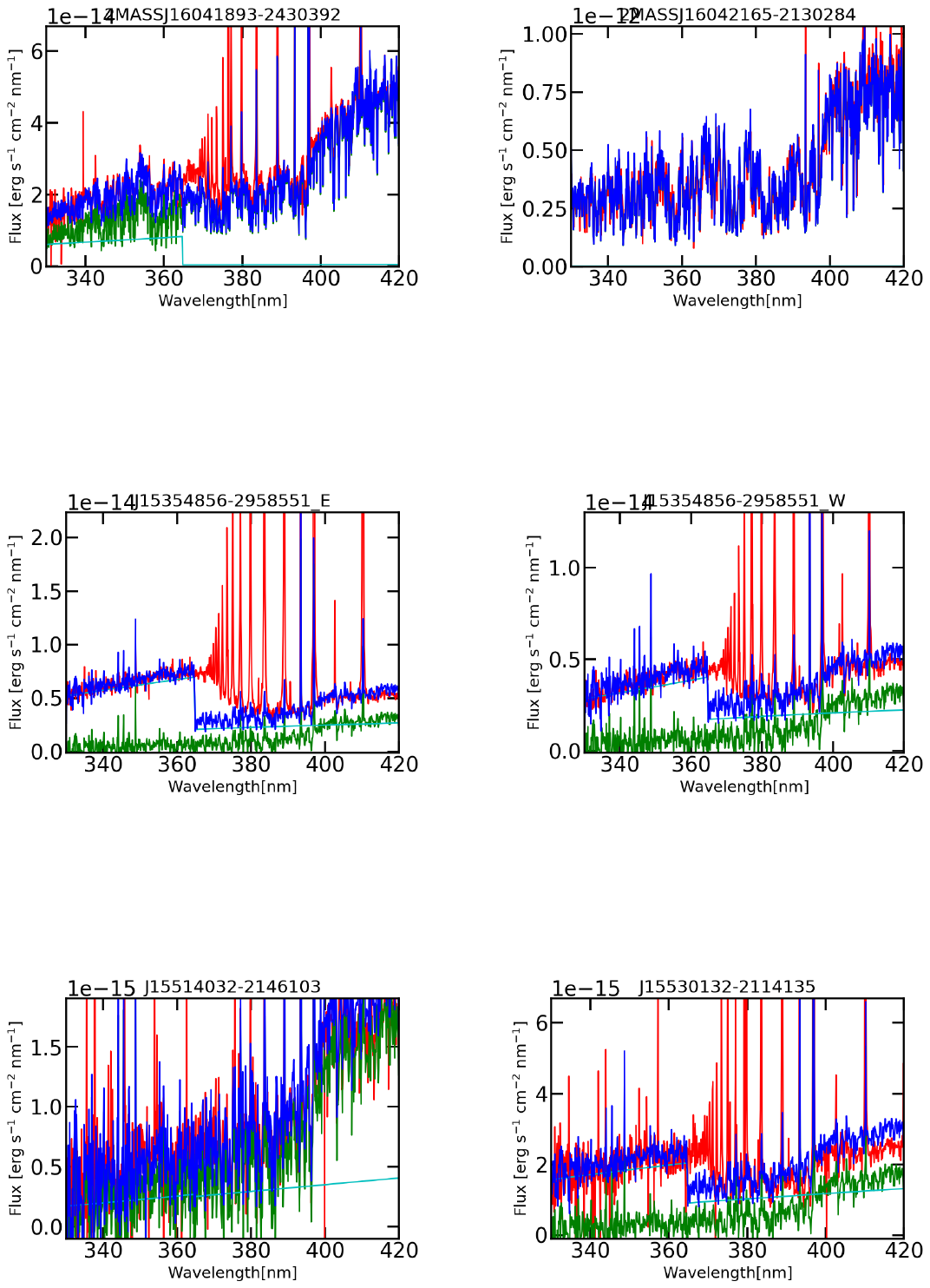}
\caption{Best fit of the continuum emission of the spectrum of the targets shown in blue together with the dereddened spectrum of the target (red). The best fit is obtained by adding the best fit photospheric template (green) and the slab model (cyan).  
     \label{fig::bj_2}}
\end{figure*}

\begin{figure*}[]
\centering
\includegraphics[width=0.9\textwidth,page=3]{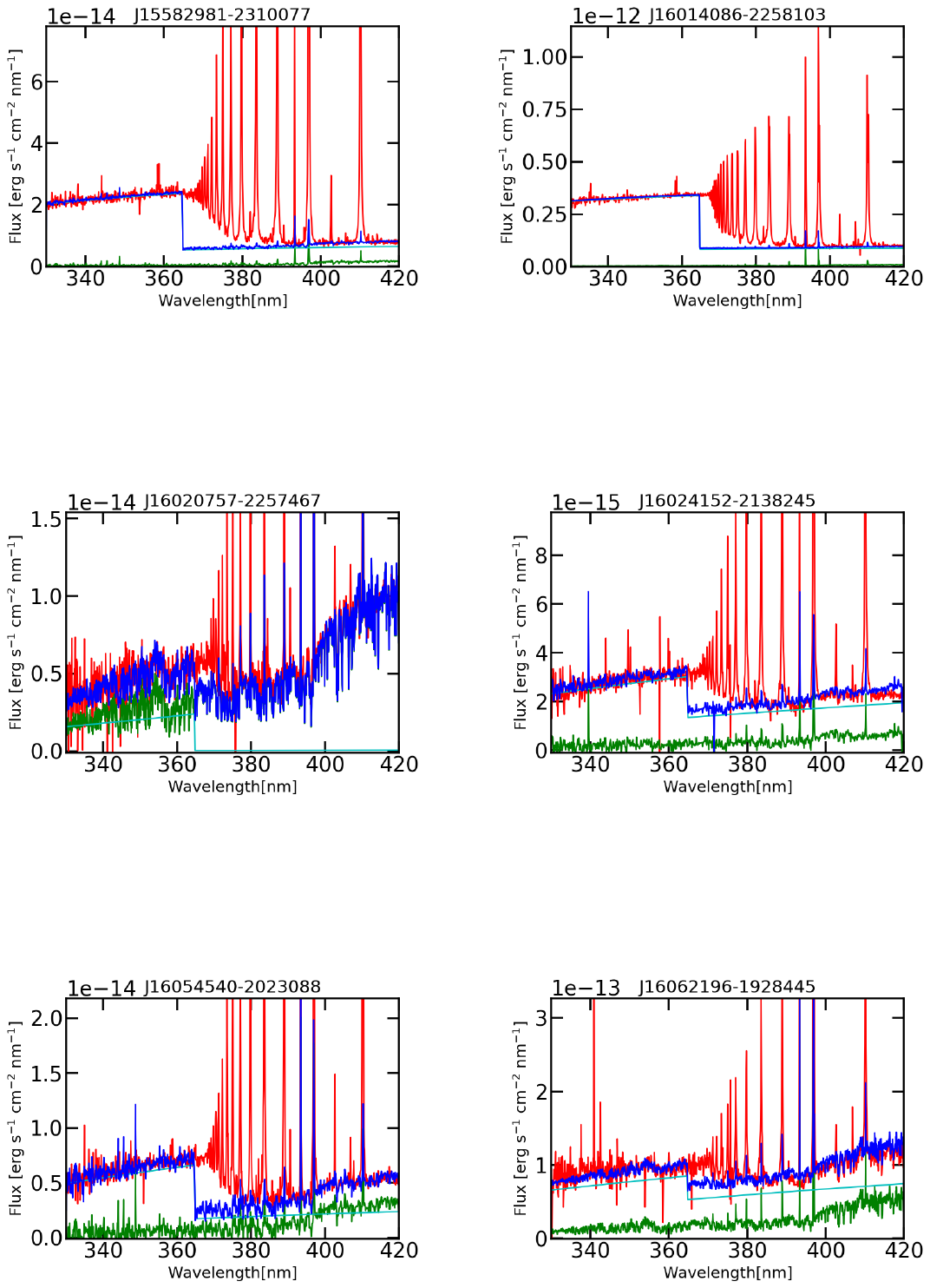}
\caption{Best fit of the continuum emission of the spectrum of the targets is shown in blue together with the dereddened spectrum of the target (red). The best fit is obtained by adding the best fit photospheric template (green) and the slab model (cyan).  
     \label{fig::bj_3}}
\end{figure*}

\begin{figure*}[]
\centering
\includegraphics[width=0.9\textwidth,page=4]{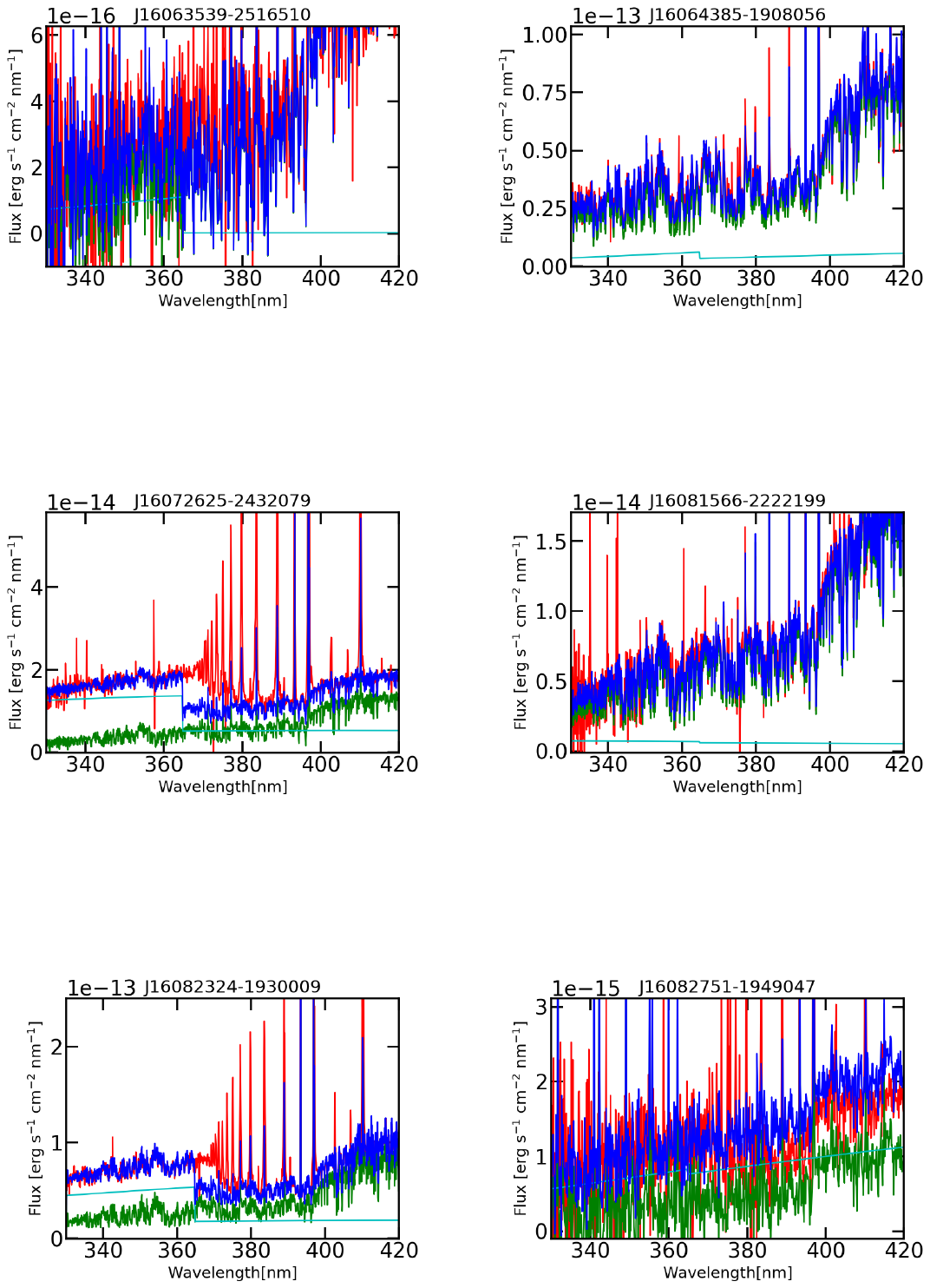}
\caption{Best fit of the continuum emission of the spectrum of the targets is shown in blue together with the dereddened spectrum of the target (red). The best fit is obtained by adding the best fit photospheric template (green) and the slab model (cyan).  
     \label{fig::bj_4}}
\end{figure*}

\begin{figure*}[]
\centering
\includegraphics[width=0.9\textwidth,page=5]{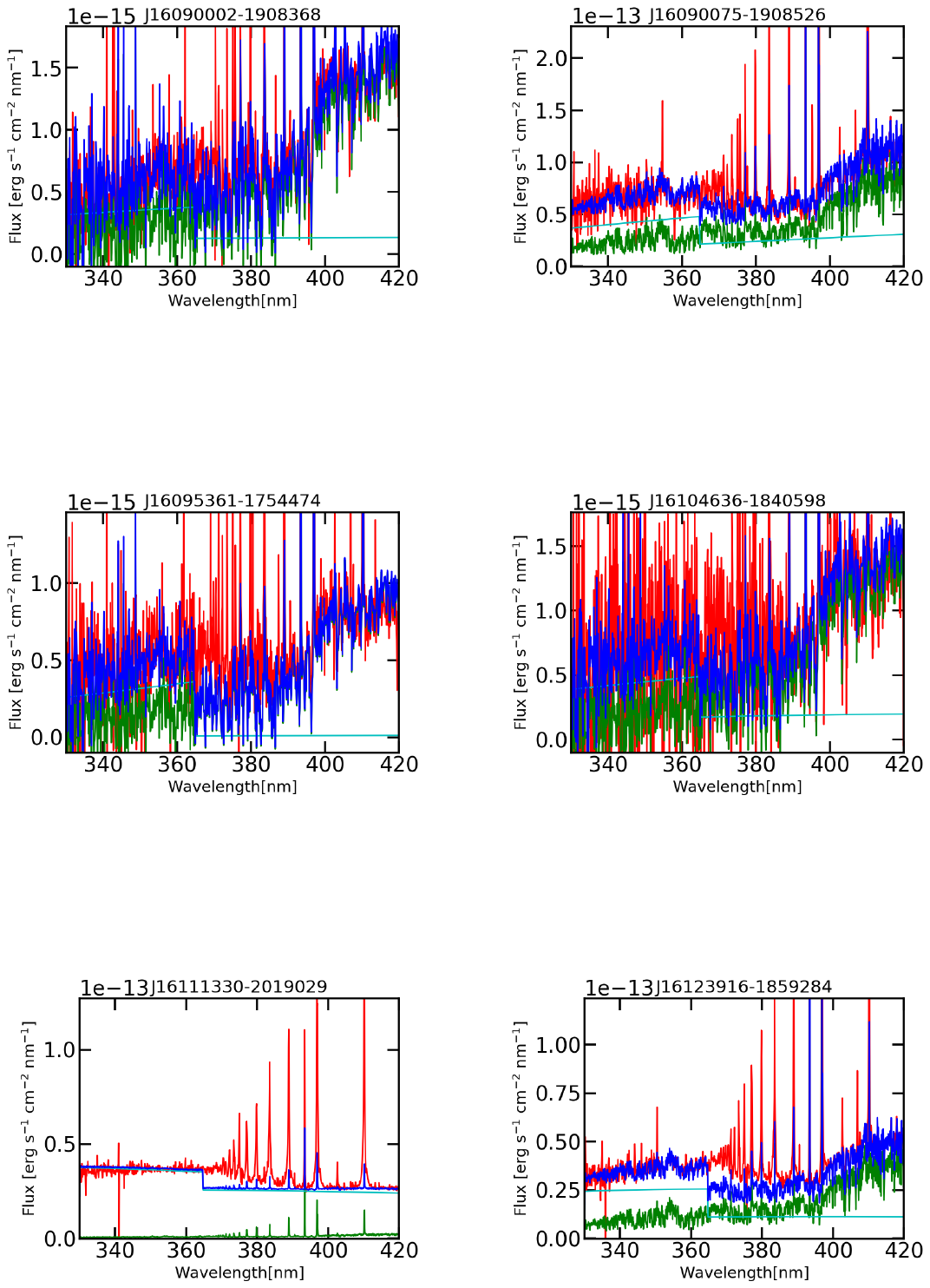}
\caption{Best fit of the continuum emission of the spectrum of the targets is shown in blue together with the dereddened spectrum of the target (red). The best fit is obtained by adding the best fit photospheric template (green) and the slab model (cyan).  
     \label{fig::bj_5}}
\end{figure*}

\begin{figure*}[]
\centering
\includegraphics[width=0.9\textwidth,page=6]{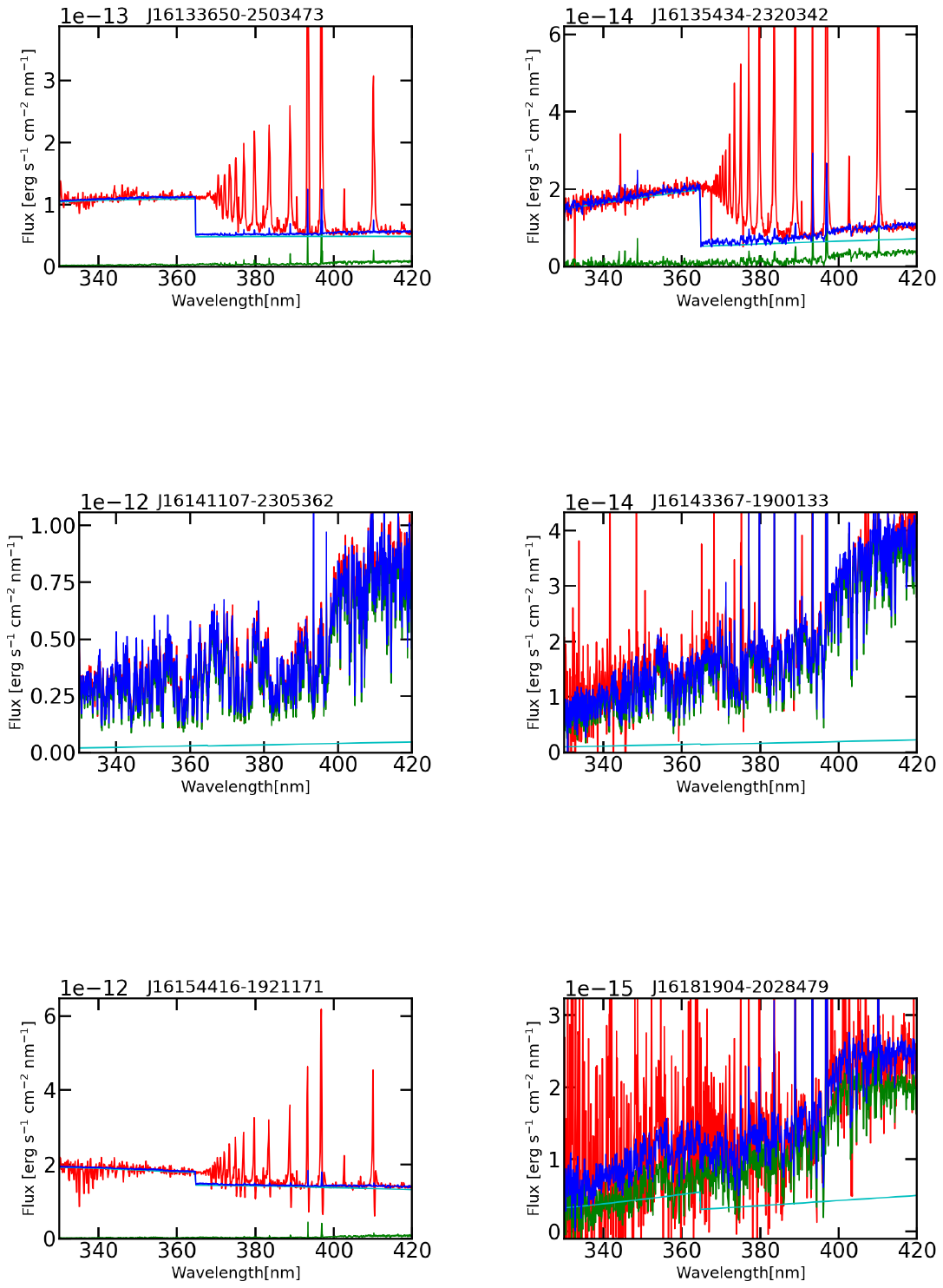}
\caption{Best fit of the continuum emission of the spectrum of the targets is shown in blue together with the dereddened spectrum of the target (red). The best fit is obtained by adding the best fit photospheric template (green) and the slab model (cyan).  
     \label{fig::bj_6}}
\end{figure*}

\subsection{Accretion luminosity vs stellar luminosity}\label{app::lacc_lstar}

\begin{figure}[]
\centering
\includegraphics[width=0.45\textwidth]{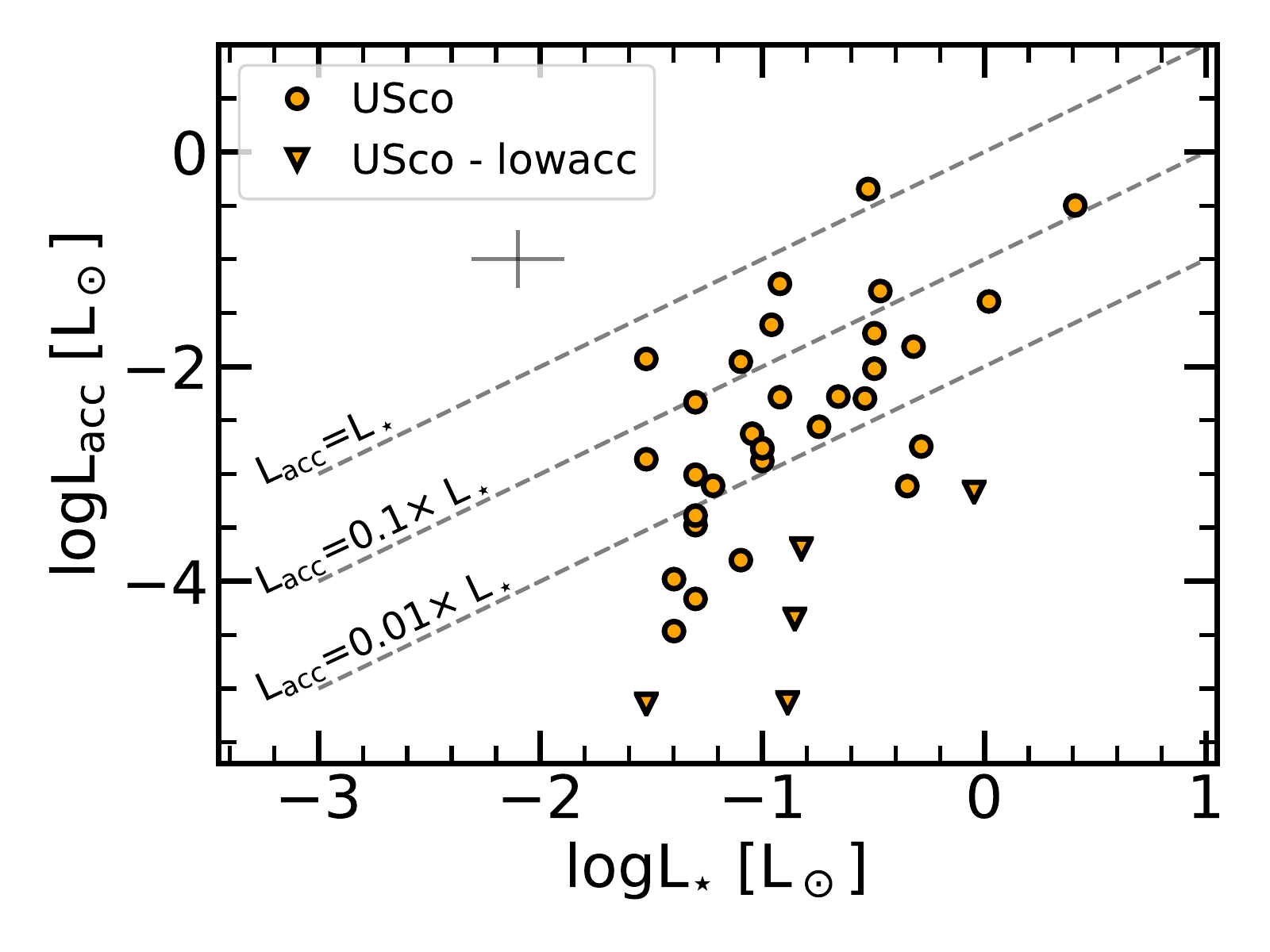}
\includegraphics[width=0.45\textwidth]{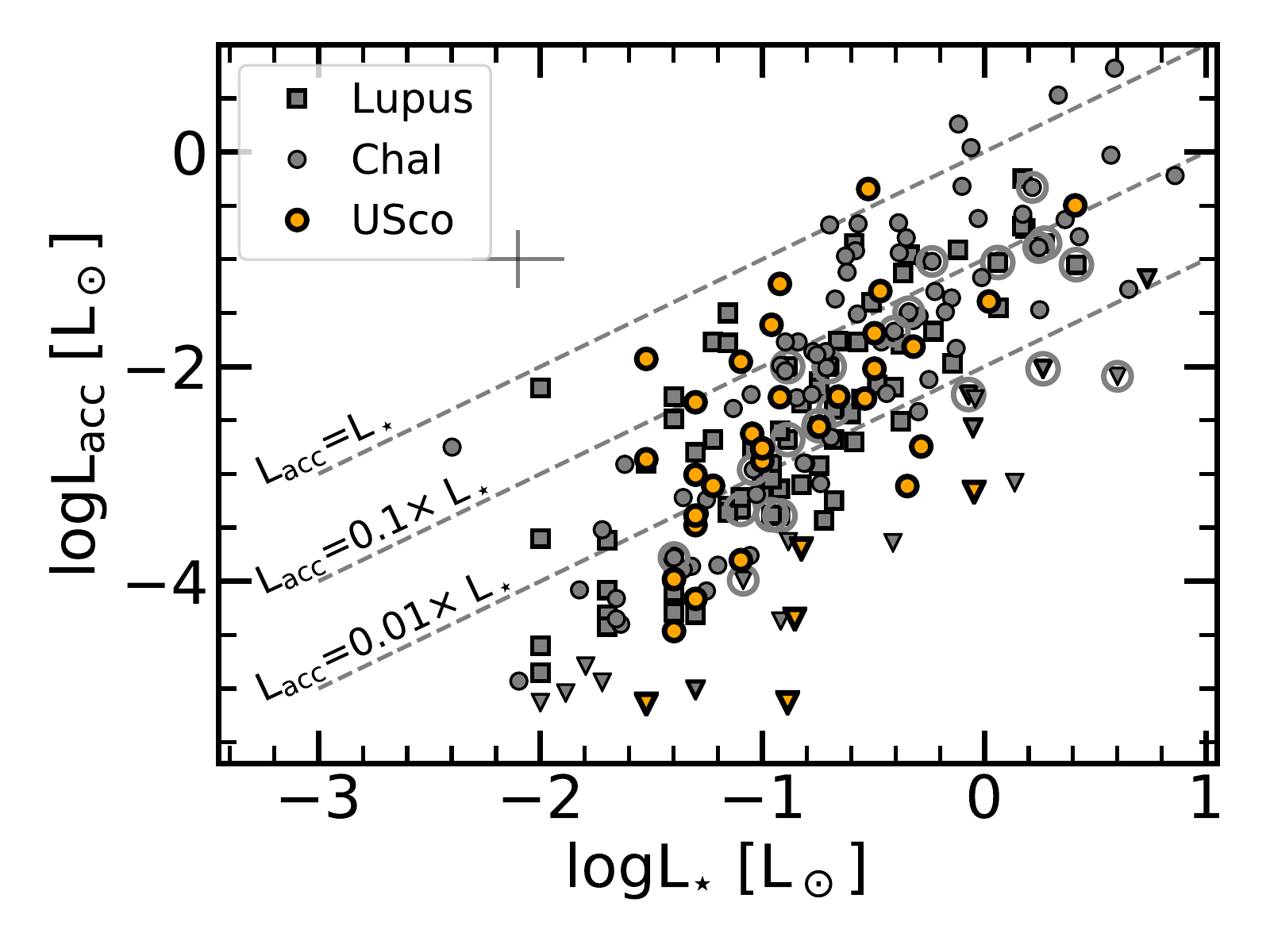}
\caption{Accretion luminosity vs stellar luminosity for the targets in the Upper Scorpius star-forming region (upper panel), and for the targets in the Upper Scorpius, Lupus, and Chamaeleon~I regions (bottom panel).
     \label{fig::lacc_lstar}}
\end{figure}

We show the relation between the accretion and stellar luminosity for the targets analyzed here and those in the star-forming regions of Lupus and Chamaeleon~I (Fig.~\ref{fig::lacc_lstar}).

\begin{thebibliography}{}

\bibitem[Alcal{\'a} et al.(2014)]{alcala14} Alcal{\'a}, J.~M., Natta, A., Manara, C.~F., et al.\ 2014, \aap, 561, A2 

\bibitem[Alcal{\'a} et al.(2017)]{alcala17} Alcal{\'a}, J.~M., Manara, C.~F., Natta, A., et al.\ 2017, \aap, 600, A20 

\bibitem[Andrews et al.(2018)]{andrews18} Andrews, S.~M., Huang, J., P{\'e}rez, L.~M., et al.\ 2018, \apjl, 869, L41

\bibitem[Ansdell et al.(2016)]{ansdell16} Ansdell, M., Williams, J.~P., van der Marel, N., et al.\ 2016, \apj, 828, 46

\bibitem[Ansdell et al.(2017)]{ansdell17} Ansdell, M., Williams, J.~P., Manara, C.~F., et al.\ 2017, \aj, 153, 240


\bibitem[Ansdell et al.(2018)]{ansdell18} Ansdell, M., Williams, J.~P., Trapman, L., et al.\ 2018, \apj, 859, 21

\bibitem[Ansdell et al.(2020)]{ansdell20} Ansdell, M., Gaidos, E., Hedges, C., et al.\ 2020, \mnras, 492, 572


\bibitem[Antoniucci et al.(2014)]{antoniucci14} Antoniucci, S., Garc{\'\i}a L{\'o}pez, R., Nisini, B., et al.\ 2014, \aap, 572, A62

\bibitem[Armitage et al.(2013)]{armitage13} Armitage, P.~J., Simon, J.~B., \& Martin, R.~G.\ 2013, \apjl, 778, L14

\bibitem[Audard et al.(2014)]{audard14} Audard, M., {\'A}brah{\'a}m, P., Dunham, M.~M., et al.\ 2014, Protostars and Planets VI, 387

\bibitem[Bai \& Stone(2013)]{bai13} Bai, X.-N., \& Stone, J.~M.\ 2013, \apj, 769, 76

\bibitem[Baraffe et al.(2015)]{B15} Baraffe, I., Homeier, D., Allard, F., \& Chabrier, G.\ 2015, \aap, 577, A42 

\bibitem[Barenfeld et al.(2016)]{barenfeld16} Barenfeld, S.~A., Carpenter, J.~M., Ricci, L., et al.\ 2016, \apj, 827, 142

\bibitem[Barenfeld et al.(2017)]{barenfeld17} Barenfeld, S.~A., Carpenter, J.~M., Sargent, A.~I., et al.\ 2017, \apj, 851, 85

\bibitem[Birnstiel et al.(2010)]{birnstiel2010} Birnstiel, T., Dullemond, C.~P., \& Brauer, F.\ 2010, \aap, 513, A79

\bibitem[Cardelli et al.(1989)]{cardelli98} Cardelli, J.~A., Clayton, G.~C., \& Mathis, J.~S.\ 1989, \apj, 345, 245

\bibitem[Carpenter et al.(2006)]{carpenter06} Carpenter, J.~M., Mamajek, E.~E., Hillenbrand, L.~A., et al.\ 2006, \apjl, 651, L49

\bibitem[Costigan et al.(2014)]{costigan14} Costigan, G., Vink, J.~S., Scholz, A., et al.\ 2014, \mnras, 440, 3444

\bibitem[Da Rio et al.(2014)]{dario14} Da Rio, N., Jeffries, R.~D., Manara, C.~F., et al.\ 2014, \mnras, 439, 3308

\bibitem[David et al.(2019)]{david19} David, T.~J., Hillenbrand, L.~A., Gillen, E., et al.\ 2019, \apj, 872, 161

\bibitem[de Albuquerque et al.(2020)]{dealb20} de Albuquerque, R.~M.~G., Gameiro, J.~F., Alencar, S.~H.~P., et al.\ 2020, arXiv e-prints, arXiv:2003.09511


\bibitem[De Marchi et al.(2017)]{demarchi17} De Marchi, G., Panagia, N., \& Beccari, G.\ 2017, \apj, 846, 110


\bibitem[Dullemond et al.(2006)]{dullemond06} Dullemond, C.~P., Natta, A., \& Testi, L.\ 2006, \apjl, 645, L69

\bibitem[Ercolano \& Pascucci(2017)]{EP17} Ercolano, B. \& Pascucci, I.\ 2017, Royal Society Open Science, 4, 170114

\bibitem[Fedele et al.(2010)]{fedele10} Fedele, D., van den Ancker, M.~E., Henning, T., et al.\ 2010, \aap, 510, A72

\bibitem[Feiden(2016)]{feiden16} Feiden, G.~A.\ 2016, \aap, 593, A99

\bibitem[Frasca et al.(2015)]{frasca15} Frasca, A., Biazzo, K., Lanzafame, A.~C., et al.\ 2015, \aap, 575, A4

\bibitem[Frasca et al.(2017)]{frasca17} Frasca, A., Biazzo, K., Alcal{\'a}, J.~M., et al.\ 2017, \aap, 602, A33

\bibitem[Freudling et al.(2013)]{reflex} Freudling, W., Romaniello, M., Bramich, D.~M., et al.\ 2013, \aap, 559, A96

\bibitem[Gaia Collaboration et al.(2016)]{gaia} Gaia Collaboration, Prusti, T., de Bruijne, J.~H.~J., et al.\ 2016, \aap, 595, A1 

\bibitem[Gaia Collaboration et al.(2018)]{gaiadr2} Gaia Collaboration, Brown, A.~G.~A., Vallenari, A., et al.\ 2018, \aap, 616, A1 

\bibitem[Haisch et al.(2001)]{haisch01} Haisch, K.~E., Lada, E.~A., \& Lada, C.~J.\ 2001, \apjl, 553, L153

\bibitem[Hartmann et al.(1998)]{hartmann98} Hartmann, L., Calvet, N., Gullbring, E., et al.\ 1998, \apj, 495, 385

\bibitem[Hartmann et al.(2016)]{HCH16} Hartmann, L., Herczeg, G., \& Calvet, N.\ 2016, \araa, 54, 135

\bibitem[Herczeg, \& Hillenbrand(2008)]{HH08} Herczeg, G.~J., \& Hillenbrand, L.~A.\ 2008, \apj, 681, 594

\bibitem[Hern{\'a}ndez et al.(2007)]{hernandez07} Hern{\'a}ndez, J., Hartmann, L., Megeath, T., et al.\ 2007, \apj, 662, 1067

\bibitem[Ingleby et al.(2014)]{ingleby14} Ingleby, L., Calvet, N., Hern{\'a}ndez, J., et al.\ 2014, \apj, 790, 47

\bibitem[Jones et al.(2012)]{jones12} Jones, M.~G., Pringle, J.~E., \& Alexander, R.~D.\ 2012, \mnras, 419, 925

\bibitem[Kausch et al.(2015)]{molecfit2} Kausch, W., Noll, S., Smette, A., et al.\ 2015, \aap, 576, A78

\bibitem[Kelly(2007)]{kelly07} Kelly, B.~C.\ 2007, \apj, 665, 1489

\bibitem[Kratter \& Lodato(2016)]{KL16} Kratter, K., \& Lodato, G.\ 2016, \araa, 54, 271

\bibitem[Lee et al.(2020)]{lee20} Lee, J., Song, I., \& Murphy, S.\ 2020, \mnras, 494, 62

\bibitem[Lodato et al.(2017)]{lodato17} Lodato, G., Scardoni, C.~E., Manara, C.~F. et al.\ 2017, \mnras, 472, 4700.

\bibitem[Luhman et al.(2003)]{luhman03} Luhman, K.~L., Stauffer, J.~R., Muench, A.~A., et al.\ 2003, \apj, 593, 1093

\bibitem[Luhman \& Mamajek(2012)]{luhman12} Luhman, K.~L., \& Mamajek, E.~E.\ 2012, \apj, 758, 31

\bibitem[Lynden-Bell \& Pringle(1974)]{LBP74} Lynden-Bell, D., \& Pringle, J.~E.\ 1974, \mnras, 168, 603

\bibitem[Mamajek et al.(2002)]{mamajek02} Mamajek, E.~E., Meyer, M.~R., \& Liebert, J.\ 2002, \aj, 124, 1670

\bibitem[Manara et al.(2013a)]{manara13a} Manara, C.~F., Testi, L., Rigliaco, E., et al.\ 2013a, \aap, 551, A107

\bibitem[Manara et al.(2013b)]{manara13b} Manara, C.~F., Beccari, G., Da Rio, N., et al.\ 2013b, \aap, 558, A114

\bibitem[Manara et al.(2014)]{manara14} Manara, C.~F., Testi, L., Natta, A., et al.\ 2014, \aap, 568, A18

\bibitem[Manara et al.(2016a)]{manara16a} Manara, C.~F., Fedele, D., Herczeg, G.~J. et al.\ 2016a, \aap, 585, A136.

\bibitem[Manara et al.(2016b)]{manara16b} Manara, C.~F., Rosotti, G., Testi, L. et al.\ 2016b, \aap, 591, L3 

\bibitem[Manara et al.(2017a)]{manara17a} Manara, C.~F., Testi, L., Herczeg, G.~J. et al.\ 2017a, \aap, 604, A127.

\bibitem[Manara et al.(2017b)]{manara17b} Manara, C.~F., Frasca, A., Alcal{\'a}, J.~M., et al.\ 2017b, \aap, 605, A86

\bibitem[Manara et al.(2018)]{manara18b} Manara, C.~F., Morbidelli, A., \& Guillot, T.\ 2018, \aap, 618, L3

\bibitem[Manara et al.(2019)]{manara19b} Manara, C.~F., Mordasini, C., Testi, L., et al.\ 2019, \aap, 631, L2

\bibitem[Modigliani et al.(2010)]{xspipe} Modigliani, A., Goldoni, P., Royer, F., et al.\ 2010, \procspie, 773728

\bibitem[Morbidelli \& Raymond(2016)]{MR16} Morbidelli, A., \& Raymond, S.~N.\ 2016, Journal of Geophysical Research (Planets), 121, 1962

\bibitem[Mulders et al.(2017)]{mulders17} Mulders, G.~D., Pascucci, I., Manara, C.~F., et al.\ 2017, \apj, 847, 31

\bibitem[Murphy et al.(2018)]{murphy18} Murphy, S.~J., Mamajek, E.~E., \& Bell, C.~P.~M.\ 2018, \mnras, 476, 3290

\bibitem[Pascucci et al.(2016)]{pascucci16} Pascucci, I., Testi, L., Herczeg, G.~J., et al.\ 2016, \apj, 831, 125

\bibitem[Pecaut, \& Mamajek(2016)]{PM16} Pecaut, M.~J., \& Mamajek, E.~E.\ 2016, \mnras, 461, 794

\bibitem[Rosotti et al.(2017)]{rosotti17} Rosotti, G.~P., Clarke, C.~J., Manara, C.~F., \& Facchini, S.\ 2017, \mnras, 468, 1631 

\bibitem[Rosotti \& Clarke(2018)]{rosotti18} Rosotti, G.~P., \& Clarke, C.~J.\ 2018, \mnras, 473, 5630 

\bibitem[Rosotti et al.(2019)]{rosotti19} Rosotti, G.~P., Tazzari, M., Booth, R.~A., et al.\ 2019, \mnras, 486, 4829

\bibitem[Rugel et al.(2018)]{rugel18} Rugel, M., Fedele, D., \& Herczeg, G.\ 2018, \aap, 609, A70

\bibitem[Sellek et al.(2020)]{sellek20} Sellek, A.~D., Booth, R.~A., \& Clarke, C.~J.\ 2020, \mnras, 492, 1279

\bibitem[Sicilia-Aguilar et al.(2010)]{sicilia-aguilar10} Sicilia-Aguilar, A., Henning, T., \& Hartmann, L.~W.\ 2010, \apj, 710, 597

\bibitem[Siess et al.(2000)]{siess00} Siess, L., Dufour, E., \& Forestini, M.\ 2000, \aap, 358, 593

\bibitem[Smette et al.(2015)]{molecfit1} Smette, A., Sana, H., Noll, S., et al.\ 2015, \aap, 576, A77

\bibitem[Soderblom et al.(2014)]{soderblom14} Soderblom, D.~R., Hillenbrand, L.~A., Jeffries, R.~D., Mamajek, E.~E., \& Naylor, T.\ 2014, Protostars and Planets VI, 219 

\bibitem[Somigliana et al.(2020)]{somigliana20} Somigliana, A., Toci, C., Lodato, G., et al.\ 2020, \mnras, 492, 1120

\bibitem[Testi et al.(2014)]{testi14} Testi, L., Birnstiel, T., Ricci, L., et al.\ 2014, Protostars and Planets VI, 339

\bibitem[Venuti et al.(2014)]{venuti14} Venuti, L., Bouvier, J., Flaccomio, E., et al.\ 2014, \aap, 570, A82

\bibitem[Venuti et al.(2019)]{venuti19} Venuti, L., Stelzer, B., Alcal{\'a}, J.~M., et al.\ 2019, \aap, 632, A46

\bibitem[Vernet et al.(2011)]{vernet11} Vernet, J., Dekker, H., D'Odorico, S., et al.\ 2011, \aap, 536, A105

\bibitem[Winter et al.(2018)]{winter18} Winter, A.~J., Clarke, C.~J., Rosotti, G., et al.\ 2018, \mnras, 478, 2700

\bibitem[Wilkinson et al.(2018)]{wilkinson18} Wilkinson, S., Mer{\'\i}n, B., \& Riviere-Marichalar, P.\ 2018, \aap, 618, A12

\bibitem[de Zeeuw et al.(1999)]{deZeeuw99} de Zeeuw, P.~T., Hoogerwerf, R., de Bruijne, J.~H.~J., et al.\ 1999, \aj, 117, 354

\bibitem[Zuckerman et al.(2014)]{zuckerman14} Zuckerman, B., Vican, L., \& Rodriguez, D.~R.\ 2014, \apj, 788, 102

\end{thebibliography}
\end{document}